\documentclass[12pt,preprint]{emulateapj-rtx4}
\usepackage{lscape}

\submitted{Accepted for Publication in The Astrophysical Journal: October 27,
2016}
\shorttitle{NGC 6273}
\shortauthors{Johnson et al.}

\newcommand\iso[2]{$^{\rm #1}$#2}

\begin{document}

\title{A CHEMICAL COMPOSITION SURVEY OF THE IRON--COMPLEX GLOBULAR
CLUSTER NGC 6273 (M 19)\footnote{Based on observations made with the NASA/ESA 
\emph{Hubble Space Telescope}, obtained at the Space Telescope Science 
Institute, which is operated by the Association of Universities for Research 
in Astronomy, Inc., under NASA contract NAS 5--26555.  These observations are 
associated with program GO--14197.  This paper includes data gathered with
the 6.5m \emph{Magellan} Telescopes located as Las Campanas Observatory, 
Chile.}}

\author{
Christian I. Johnson\altaffilmark{1,2},
Nelson Caldwell\altaffilmark{1},
R. Michael Rich\altaffilmark{3},
Mario Mateo\altaffilmark{4},
John I. Bailey, III\altaffilmark{5},
William I. Clarkson\altaffilmark{6},
Edward W. Olszewski\altaffilmark{7}, and
Matthew G. Walker\altaffilmark{8}
}

\altaffiltext{1}{Harvard--Smithsonian Center for Astrophysics, 60 Garden
Street, MS--15, Cambridge, MA 02138, USA; cjohnson@cfa.harvard.edu; 
ncaldwell@cfa.harvard.edu}

\altaffiltext{2}{Clay Fellow}

\altaffiltext{3}{Department of Physics and Astronomy, UCLA, 430 Portola Plaza,
Box 951547, Los Angeles, CA 90095-1547, USA; rmr@astro.ucla.edu}

\altaffiltext{4}{Department of Astronomy, University of Michigan, Ann Arbor,
MI 48109, USA; mmateo@umich.edu}

\altaffiltext{5}{Leiden Observatory, Leiden University, P. O. Box 9513, 2300RA 
Leiden, The Netherlands; baileyji@strw.leidenuniv.nl}

\altaffiltext{6}{Department of Natural Sciences, University of 
Michigan--Dearborn, 4901 Evergreen Road, Dearborn, MI 48128, USA; 
wiclarks@umich.edu}

\altaffiltext{7}{Steward Observatory, The University of Arizona, 933 N. Cherry
Avenue, Tucson, AZ 85721, USA; eolszewski@as.arizona.edu}

\altaffiltext{8}{McWilliams Center for Cosmology, Department of Physics, 
Carnegie Mellon University, 5000 Forbes Avenue, Pittsburgh, PA 15213, USA; 
mgwalker@andrew.cmu.edu}

\begin{abstract}

Recent observations have shown that a growing number of the most massive 
Galactic globular clusters contain multiple populations of stars with different
[Fe/H] and neutron--capture element abundances.  NGC 6273 has only recently 
been recognized as a member of this ``iron--complex" cluster class, and we 
provide here a chemical and kinematic analysis of $>$ 300 red giant branch 
(RGB) and asymptotic giant branch (AGB) member stars using high resolution 
spectra obtained with the \emph{Magellan}--M2FS and \emph{VLT}--FLAMES 
instruments.  Multiple lines of evidence indicate that NGC 6273 possesses an 
intrinsic metallicity spread that ranges from about [Fe/H] = --2 to --1 dex, 
and may include at least three populations with different [Fe/H] values.  
The three populations identified here contain separate first (Na/Al--poor) and 
second (Na/Al--rich) generation stars, but a Mg--Al anti--correlation may only 
be present in stars with [Fe/H] $\ga$ --1.65.  The strong correlation between 
[La/Eu] and [Fe/H] suggests that the s--process must have dominated the heavy 
element enrichment at higher metallicities.  A small group of stars with low 
[$\alpha$/Fe] is identified and may have been accreted from a former 
surrounding field star population.  The cluster's large abundance variations
are coupled with a complex, extended, and multimodal blue horizontal branch 
(HB).  The HB morphology and chemical abundances suggest that NGC 6273 
may have an origin that is similar to $\omega$ Cen and M 54.

\end{abstract}


\section{INTRODUCTION}

Galactic globular clusters are no longer considered pure simple stellar 
populations.  Although large and often (anti--)correlated star--to--star light 
element abundance variations have long been known to exist within individual 
globular clusters (e.g., Cohen 1978; Peterson 1980; Cottrell \& Da Costa 1981;
Sneden et al. 1991; Pilachowski et al. 1996a; Kraft et al. 1997; Shetrone \& 
Keane 2000; Gratton et al. 2001; Ivans et al. 2001), the ubiquitous nature of 
their peculiar chemical compositions has only recently been recognized.  Large 
sample spectroscopic surveys have revealed that all but perhaps the lowest mass
clusters (Walker et al. 2011; Villanova et al. 2013; Salinas \& Strader 2015) 
exhibit similar, but not identical, (anti--)correlations among elements 
ranging from carbon to aluminum (e.g., Carretta et al. 2009a, 2009b; 
M{\'e}sz{\'a}ros et al. 2015).  In many cases, He enhancements coincide with 
increased abundances of N, Na, and Al and decreased abundances
of C, O, and Mg (e.g., Bragaglia et al. 2010a, 2010b; Dupree et al. 2011; 
Pasquini et al. 2011; Villanova et al. 2012; Marino et al. 2014a; Mucciarelli
et al. 2014).  Except for CN variations due to \emph{in situ} mixing, these 
interconnected light element abundance patterns may be unique to old ($\ga$ 6
Gyr) globular cluster environments (e.g., Pilachowski et al. 1996b; Sneden et 
al. 2004; Mucciarelli et al. 2008; Bragaglia et al. 2014).

Large light element abundance variations can have a significant effect on a 
star's structure and spectrum (e.g., see Piotto et al. 2015; their Figure 1), 
and recent near UV observations from the \emph{Hubble Space Telescope} 
(\emph{HST}) have exploited this property to reveal a further connection 
between chemical compositions and globular cluster formation.  A key 
observational constraint for globular cluster formation scenarios is whether 
the range of light element abundances follows a continuous distribution or 
falls into discrete groups.  Although some purely spectroscopic evidence 
supports clusters hosting discrete groups with unique light element chemistry 
(e.g., Carretta et al. 2009a, 2014; Johnson \& Pilachowski 2010; Carretta 2014,
2015; Cordero et al. 2014; Roederer \& Thompson 2015), \emph{HST} photometry
has been particularly efficient at showing that most or all Galactic globular
clusters host multiple distinct populations rather than continuous 
distributions (e.g., Piotto et al. 2007, 2015; Bragaglia et al. 2010a; Milone 
et al. 2013, 2015a, 2015b; Marino et al. 2016).  The combined data from 
spectroscopy and photometry provide strong evidence that globular clusters 
experienced multiple rounds of star formation.  However, the detailed processes
by which globular clusters form, and the nucleosynthetic origins of the light 
element abundance variations, remain unresolved issues (e.g., see recent 
discussions in Valcarce \& Catelan 2011; Bastian et al. 2015; Bastian \& Lardo 
2015; Renzini et al. 2015; D'Antona et al. 2016)

Despite most globular clusters exhibiting large light element abundance 
variations, most systems do not display the same complexity for the 
heavier elements.  The [Fe/H]\footnote{[A/B]$\equiv$log(N$_{\rm A}$/N$_{\rm B}$)$_{\rm star}$--log(N$_{\rm A}$/N$_{\rm B}$)$_{\sun}$ and log $\epsilon$(A)$\equiv$log(N$_{\rm A}$/N$_{\rm H}$)+12.0 for elements A and B.} and [X/Fe] 
ratios for most $\alpha$ and Fe--peak elements vary by $\sim$0.1 dex or less
within an individual cluster (e.g., Carretta et al. 2009c), but intrinsic 
variations at the few percent level may be present for all elements 
(Yong et al. 2013).  Some clusters exhibit primordial abundance variations 
for elements produced by the rapid neutron--capture process (r--process), but 
many do not (e.g., Roederer 2011).  Most clusters also fail to show chemical 
signatures of extended star formation histories, such as elevated slow
neutron--capture (s--process) abundances or low [$\alpha$/Fe] ratios.  More
metal--rich clusters tend to exhibit stronger s--process signatures 
(e.g., higher average [Ba/Eu] or [La/Eu] ratios) than their more metal--poor 
counterparts (e.g., Simmerer et al. 2003; Gratton et al. 2004; James et al. 
2004; Cohen \& Mel{\'e}ndez 2005; Carretta et al. 2007; D'Orazi et al. 2010; 
Worley \& Cottrell 2010), but these differences are likely driven by the 
broader chemical enrichment of the Galaxy.

Interestingly, a growing number of clusters have been discovered that exhibit
chemical and morphological characteristics consistent with extended star 
formation histories, and may represent a new class of objects.  These 
``iron--complex"\footnote{Note that iron--complex clusters are the same as the 
``anomalous" and ``s--Fe--anomalous" clusters discussed in Marino et al. 
(2015).  As mentioned in Johnson et al. (2015a), we prefer to avoid using the 
word ``anomalous" in this context because the word has multiple historical 
definitions.  Additionally, the anomalous label may not be appropriate if 
additional systems continue to be found.} clusters are characterized as 
having: (1) broadened or multimodal [Fe/H] distribution functions with 
dispersions exceeding $\sim$0.1 dex when measured using high resolution 
spectra\footnote{Note that the metallicity dispersions are contested for some
clusters (Mucciarelli et al. 2014; Lardo et al. 2016; but see also Lee 2016).}; (2) complex color--magnitude diagrams and split RGB sequences when 
observed with \emph{hk} narrow band photometry (e.g., Lee 2015; Lim et al. 
2015); (3) and correlated abundances of [Fe/H] and elements likely produced by 
the main s--process (e.g., Ba and La).  To date, $\sim$10 iron--complex 
clusters have been discovered (e.g., see Da Costa 2016a, their Table 1; Marino 
et al. 2015, their Table 10)\footnote{Terzan 5 is not included in the 
aforementioned lists but has also been shown to contain multiple generations 
of stars with distinct chemical compositions (Ferraro et al. 2009; Origlia et 
al. 2011, 2013; Massari et al. 2014).}.  Many of these systems also have about 
the same metallicity ([Fe/H] $\sim$ --1.7), have very blue and extended 
horizontal branch (HB) morphologies, and are among the most massive clusters 
in the Galaxy (M$_{\rm V}$ $\la$ --8).  The iron--complex cluster M 54 may be 
the nuclear star cluster of the Sagittarius dwarf galaxy (e.g., Bellazzini et 
al. 2008), and the most massive iron--complex cluster omega Centauri 
($\omega$ Cen) is strongly suspected to be a stripped dwarf galaxy nucleus as 
well (e.g., Bekki \& Freeman 2003).  Similarly, the iron--complex clusters NGC 
1851 and M 2 may also be the stripped cores of former dwarf galaxies 
(e.g., Olszewski et al. 2009; Kuzma et al. 2016).  Therefore, iron--complex 
clusters may be the relics of more massive systems, the remnants of previous 
Milky Way accretion events, and/or trace a particular time or accretion period 
in the Galaxy's formation history.

Among the iron--complex cluster class, $\omega$ Cen, M 54 and the Sagittarius 
system, M 2, NGC 5286, and NGC 6273 (M 19) stand out as particularly 
interesting.  These clusters exhibit broad metallicity distributions with 
discrete populations occurring near the same [Fe/H] values, and also host trace
populations of metal--rich stars with peculiar chemical compositions (e.g., 
Pancino et al. 2002; Carretta et al. 2010a; Johnson \& Pilachowski 2010; Marino 
et al. 2011a, 2015; McWilliam et al. 2013; Yong et al. 2014; Johnson et al. 
2015a).  In order to investigate this phenomenon further, we have obtained high 
resolution spectra of $>$ 800 red giant branch (RGB) and asymptotic giant branch
(AGB) stars located near the massive bulge cluster NGC 6273.  Following Johnson
et al. (2015a), Han et al. (2015), and Yong et al. (2016), we aim to 
investigate the cluster's metallicity distribution function and trace the 
cluster's detailed chemical composition across its various stellar populations.

\section{OBSERVATIONS AND DATA REDUCTION}

\subsection{\emph{Magellan} Spectroscopic Data}

\begin{figure}
\epsscale{1.00}
\plotone{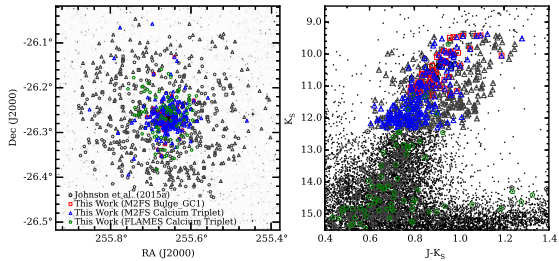}
\caption{\emph{Left:} the sky coordinates of all targets observed for this
work and Johnson et al. (2015a) are superimposed on a 2MASS (Skrutskie et
al. 2006) J--band image centered on NGC 6273.  The black, red, blue, and green
symbols indicate stars that are radial velocity members, and the grey symbols
indicate stars that are likely not cluster members.  \emph{Right:} a 2MASS
J--K$_{\rm S}$ color--magnitude diagram is shown with the NGC 6273 member and
non--member stars indicated using the same symbol and color designations as in
the left panel.}
\label{f1}
\end{figure}

In Johnson et al. (2015a), we identified an intrinsic metallicity spread in
NGC 6273, and noted the existence of several stars redder than the formal RGB 
that could belong to an even more metal--rich component.  Since the previous
observations were restricted to the color range 0.7 $\le$ J--K$_{\rm S}$ $\le$ 
1.0 on the upper RGB, we expanded the target selection criteria for the new 
observations to include stars in the color range 0.6 $\le$ J--K$_{\rm S}$ $\le$
1.3.  The new observations also span luminosities from the HB to the RGB--tip, 
and range from 0.53--13.98$\arcmin$ in projected distance from the cluster 
center (see Figure \ref{f1}).  However, stars closer to the cluster center 
were given higher priorities in the target ranking process.  All coordinates 
and photometry for the target selection process were taken from the Two Micron 
All Sky Survey (2MASS; Skrutskie et al. 2006) database.

In order to efficiently obtain a large number of high resolution spectra, we
employed the Michigan/\emph{Magellan} Fiber System (M2FS; Mateo et al. 2012)
and MSpec multi--object spectrograph mounted on the \emph{Magellan}--Clay 6.5m 
telescope.  In single order mode, M2FS is capable of placing 256 1.2$\arcsec$
fibers on targets across a nearly 30$\arcmin$ field--of--view.  However, 
additional orders can be observed simultaneously using a cross--disperser, at 
the expense of fewer targets.  We utilized both options for this project.  The
first setup operated in single order mode and was optimized to observe the 
8542 \AA\ and 8662 \AA\ near--infrared \ion{Calcium}{2} Triplet (CaT) lines.  
These data provided radial velocities and CaT metallicities for 466 stars, and 
permitted an investigation into the full spatial, color, and metallicity extent 
of NGC 6273.  The second setup (``Bulge$\_$GC1" filter) included 6 consecutive 
orders, spanned 6120--6720 \AA, allowed for up to 48 fibers to be allocated 
per configuration, and was used to obtain radial velocities and detailed 
chemical abundances for 82 stars.  As can be seen in Figure \ref{f1}, both 
data sets spanned broad color and radial distance ranges, but the CaT data 
extended to fainter stars.  

Both instrument setups utilized a four amplifier slow readout mode and were 
binned 2 $\times$ 1 (spatial $\times$ dispersion).  The CaT and Bulge$\_$GC1 
observations were taken with the 180$\mu$m (widest) and 125$\mu$m slits, 
respectively.  However, both setups yielded approximately the same resolving 
power of R $\equiv$ $\lambda$/$\Delta$$\lambda$ $\approx$ 27,000, based on an 
examination of the ThAr wavelength calibration spectra.  The two CaT fields
were observed for a total of 10,200 seconds, and the two Bulge$\_$GC1 fields
were observed for a total of 21,600 seconds.  A summary of the observation 
dates, instrument configurations, and integration times is provided in Table 1.

For data reduction, we followed the procedures outlined in Johnson et al. 
(2015b; see their Section 2.3).  Briefly, we used standard IRAF\footnote{IRAF 
is distributed by the National Optical Astronomy Observatory, which is operated
by the Association of Universities for Research in Astronomy, Inc., under 
cooperative agreement with the National Science Foundation.} tasks to apply
the bias correction, trim the overscan regions, correct for dark current, 
and combine the individual amplifier images from each CCD into single
images.  The IRAF \emph{dohydra} task was used for aperture identification
and tracing, flat--field correction, scattered light removal, wavelength
calibration, cosmic ray removal, and spectrum extraction.  For the CaT data,
we did not apply any corrections for fringing beyond the flat--field 
correction.  A master sky spectrum was created for each exposure by combining 
the individual sky fiber spectra.  The target spectra were then sky corrected 
using the \emph{skysub} routine.  Finally, the individual extracted spectra 
for each star were co--added separately, normalized with the \emph{continuum} 
routine, and corrected for telluric absorption lines using the \emph{telluric} 
task.  Typical signal--to--noise (S/N) ratios ranged from about 20--100 per 
pixel for the CaT data and 30--100 per pixel for the Bulge$\_$GC1 data.

\subsection{\emph{Very Large Telescope} Spectroscopic Data}

We supplemented the M2FS CaT data set with additional observations of 300 RGB 
stars taken with the \emph{Very Large Telescope} (\emph{VLT}) FLAMES--GIRAFFE 
instrument.  The data were downloaded from the European Southern Observatory 
(ESO) Science Archive Facility under request number 210062\footnote{Based on 
observations made with ESO Telescopes at the La Silla Paranal Observatory 
under program ID 093.D--0628.}.  The FLAMES observations spanned a broad 
range of magnitudes, but were generally fainter than the M2FS data.  However, 
the spatial coverage between the two data sets was similar 
(see Figure \ref{f1}).  Note that we have only included stars for which we 
could identify a 2MASS source within 2$\arcsec$ of the coordinates provided in 
the image headers.

All of the FLAMES--GIRAFFE observations were obtained using the HR21 setup, 
which provides R $\approx$ 18,000 spectra from 8482--9000 \AA.  However, we 
only analyzed the region spanning 8500--8700 \AA, which is similar to the 
M2FS--CaT data and includes the same 8542 \AA\ and 8662 \AA\ CaT features.
The observations were taken via six configurations, each with an integration 
time of 2445 seconds.  Most stars were observed in two configurations, but not
always with the same fiber each time.  A small number of stars were observed 
in three or more configurations, and a few were observed only once.  A 
summary of the observation dates for each configuration is provided in Table 1.

The data were primarily reduced using the GIRAFFE Base--Line Data Reduction 
Software (girBLDRS\footnote{The girBLDRS software can be downloaded at: 
http://girbldrs.sourceforge.net/.}) package.  The girBLDRS suite was used to 
carry out basic CCD processing tasks (e.g., bias correction and overscan 
trimming) and also the more advanced multi--fiber tasks we performed with
\emph{dohydra} for the M2FS data (see Section 2.1).  Similar to the M2FS CaT
data, we did not apply any further corrections for fringing beyond the 
flat--field correction.  The sky subtraction, continuum normalization, and 
spectrum combining were carried out with the same IRAF routines as used for 
the M2FS data.  However, since the FLAMES data were obtained over the course 
of several weeks to months, we applied the heliocentric velocity corrections 
provided in the image headers before combining the multiple exposures.  The 
final S/N values are comparable to those of the M2FS CaT data.

\subsection{\emph{HST} Imaging Data}

NGC 6273 is known to have a broad RGB and a peculiar HB morphology that is 
similar to $\omega$ Cen (Piotto et al. 1999; Momany et al. 2004; Brown et al. 
2010, 2016; Han et al. 2015).  Therefore, in support of our spectroscopic 
observations we have obtained new \emph{HST} Wide Field Camera 3 UVIS channel 
(WFC3/UVIS) data centered on NGC 6273 that includes the F336W, F438W, F555W, 
and F814W filters.  The observations were split into a series of short and long
exposures, taken over the course of 4 orbits, that ranged in duration from 
10--685 seconds.  A post--flash of 2.0--4.7 seconds was included for all 
exposures, and the \emph{BLADE = A} option was set for all of the 10 second 
exposures to minimize shutter--induced vibration (see Section 6.11.4 of the 
WFC3 handbook\footnote{The WFC3 handbook is available at:
http://www.stsci.edu/hst/wfc3/documents/handbooks/currentIHB/.}).  A summary of
the filter choices, integration times, and observation dates is provided in
Table 1.

The basic data reductions were carried out by the Space Telescope Science
Institute's WFC3 pipeline, but we only performed analyses on the CTE--corrected
\emph{flc} images.  All photometry was obtained using the 
DOLPHOT\footnote{DOLPHOT can be downloaded at:
http://americano.dolphinsim.com/dolphot/.} (Dolphin 2000) package and its
associated WFC3 module.  The DOLPHOT parameters closely followed the values 
recommended by Williams et al. (2014) and provided by the DOLPHOT/WFC3 
documentation for point sources in crowded fields.  No special attempt was made
to recover saturated stars; however, only a small number of the brightest 
stars, predominantly in the F814W filter, were lost due to saturation.

\begin{figure}
\epsscale{1.00}
\plotone{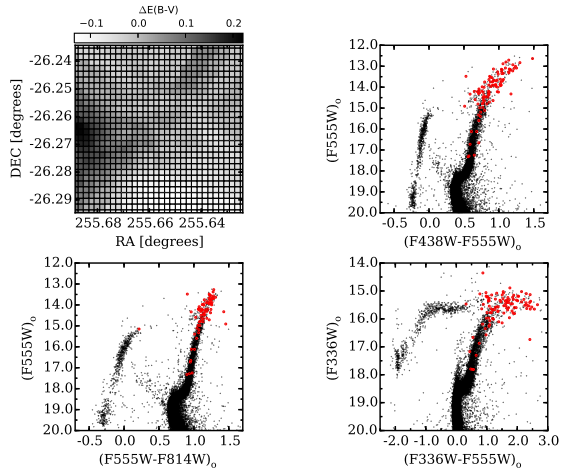}
\caption{The top left panel illustrates the spatial variations in differential
reddening, $\Delta$E(B--V), across the WFC3 field of NGC 6273, and is in good
agreement with the map provided by Alonso--Garc{\'{\i}}a et al. (2012).  Note
that the high reddening region on the eastern side of the cluster core
correlates with the known position of an interstellar cloud (e.g., Harris et
al. 1976).  The remaining panels show dereddened color--magnitude diagrams
using combinations of the F336W, F438W, F555W, and F814W filters.  The open red
circles indicate stars from our sample and Johnson et al. (2015) that have
radial velocities consistent with cluster membership.  All WFC3 photometry is
on the VEGAMAG system.}
\label{f2}
\end{figure}

As noted by several previous authors (Racine 1973; Harris et al. 1976; Piotto 
et al. 1999; Davidge 2000; Valenti et al. 2007; Brown et al. 2010; 
Alonso--Garc{\'{\i}}a et al. 2012), differential reddening is a significant
concern along lines--of--sight near NGC 6273.  Previous work estimated that the
cluster has E(B--V) = 0.31--0.47 magnitudes and $\Delta$E(B--V) $\sim$ 0.2--0.3
magnitudes.  We observe a similar reddening range of $\Delta$E(B--V) = 0.36 
magnitudes using corrections kindly provided by A. Milone (2016, private
communication; see also Milone et al. 2012 for an outline of the dereddening 
procedure) via the F336W and F814W data sets.  Additionally, we find that 
adopting an absolute color excess of E(B--V) = 0.37 magnitudes places the 
coolest HB stars at approximately the correct F555W magnitude, assuming a 
distance of 9 kpc (Piotto et al. 1999)\footnote{Note that we have adopted the 
extinction coefficients provided by Girardi et al. (2008) and updated at:
http://stev.oapd.inaf.it/cgi-bin/cmd, for all filters.  We have also employed
a ``standard" extinction curve with A$_{\rm V}$ = 3.1E(B--V).  However, see
Udalski (2003), Gosling et al. (2009), and Nataf et al. (2013, 2016) for
discussions regarding the validity of adopting a standard extinction
curve near the Galactic center.}.  Further details regarding the photometric 
analysis, including the dereddening procedure, will be provided in a future 
publication.  However, in Figure \ref{f2} we show the smoothed reddening map 
of the WFC3 field, and include several dereddened color--magnitude diagrams 
with the radial velocity members identified.

\section{RADIAL VELOCITIES AND CLUSTER MEMBERSHIP}

Radial velocities were measured for all M2FS and FLAMES spectra using the 
XCSAO (Kurtz \& Mink 1998) cross--correlation code.  The velocities were
measured relative to a synthetic stellar spectrum of an evolved RGB star with
[Fe/H] = --1.60, which is approximately the average metallicity of NGC 6273 
(Johnson et al. 2015a).  The template spectrum was smoothed and rebinned to 
match the resolution and sampling of the observed spectra.  Heliocentric
velocity corrections were calculated with IRAF's \emph{rvcorrect} utility for
the M2FS data, and for the FLAMES data we used the corrections provided in the
image headers.  The heliocentric corrections were applied to all of the 
spectra before being measured with XCSAO.

\begin{figure}
\epsscale{1.00}
\plotone{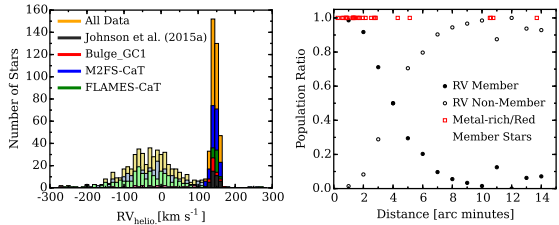}
\caption{\emph{Left:} a radial velocity histogram is shown for all of the
spectroscopic data sets used here.  Stars with heliocentric radial velocities
between $+$120 and $+$170 km s$^{\rm -1}$ were considered cluster members, and
are indicated by the dark colored histograms.  The light colored histograms
show the radial velocity distributions of the non--members.  The data are
sampled in 10 km s$^{\rm -1}$ bins.  \emph{Right:} a plot of the
member/non--member ratio as a function of the projected distance from the
cluster center.  Cluster membership was assigned using a star's heliocentric
radial velocity.  The open red boxes indicate the projected radial distances
for radial velocity member stars with [Fe/H] $>$ --1.35 and/or that lie redward
of the dominant RGBs seen in Figures \ref{f1} and \ref{f2}.}
\label{f3}
\end{figure}

For the Bulge$\_$GC1 spectra, we measured the velocities using the 
6140--6270 \AA\ window because it contains several lines suitable for 
cross--correlation but avoids very broad lines (e.g., H$\alpha$) and any
residual telluric features.  For the M2FS and FLAMES CaT data, we used the 
full spectral window from 8500--8700 \AA, but avoided the strong CaT lines.
A histogram of the heliocentric radial velocity (RV$_{\rm helio.}$) 
distributions for each data set, including data from Johnson et al. (2015a), 
is shown in Figure \ref{f3}.  Using these data, we considered stars with 
RV$_{\rm helio.}$ between $+$120 and $+$170 km s$^{\rm -1}$ to be cluster 
members.  Therefore, the new Bulge$\_$GC1, M2FS CaT, and FLAMES CaT data 
provided average velocities and dispersions of $+$143.15 km s$^{\rm -1}$ 
($\sigma$ = 9.53 km s$^{\rm -1}$), $+$144.74 km s$^{\rm -1}$ ($\sigma$ = 8.79 
km s$^{\rm -1}$), and $+$145.76 km s$^{\rm -1}$ ($\sigma$ = 7.12 km 
s$^{\rm -1}$), respectively, for the cluster members.  Similarly, the average 
RV$_{\rm helio.}$ value for the combined data sets is $+$144.71 km s$^{\rm -1}$ 
($\sigma$ = 8.57 km s$^{\rm -1}$), which is in good agreement with recent 
measurements (Johnson et al. 2015a; Yong et al. 2016).  For the non--member
stars, we found the average velocity and dispersion to be --29.36 km 
s$^{\rm -1}$ and $\sigma$ = 77.02 km s$^{\rm -1}$.  These values are in 
agreement with previous kinematic observations of similar off--axis bulge 
fields (e.g., Kunder et al. 2012; Ness et al. 2013a; Zoccali et al. 2014).

The average RV$_{\rm helio.}$ uncertainties are 0.31 km s$^{\rm -1}$ 
($\sigma$ = 0.27 km s$^{\rm -1}$), 1.09 km s$^{\rm -1}$ 
($\sigma$ = 0.69 km s$^{\rm -1}$), and 0.88 km s$^{\rm -1}$ 
($\sigma$ = 0.06 km s$^{\rm -1}$) for the Bulge$\_$GC1, M2FS CaT, and FLAMES 
CaT data, respectively.  These values represent the measurement uncertainties
from the XCSAO cross--correlation routine.  However, 57 stars were observed in
at least two different setups, including the data from Johnson et al. (2015a),
and we measured an average dispersion between repeat measurements of 
1.31 km s$^{\rm -1}$.  If we ignore the four outliers\footnote{Note that we 
have not rejected the outlier stars from the list of member stars nor the 
chemical abundance analysis.} with dispersions $>$ 5 km s$^{\rm -1}$, the 
average dispersion decreases to 0.88 km s$^{\rm -1}$.  Therefore, we regard 
$\sim$1 km s$^{\rm -1}$ as a reasonable estimate of the systematic uncertainty 
due to the use of different instruments, configurations, and wavelength regions.

As can be seen in Figure \ref{f3}, the systemic cluster velocity is well
separated from the broad field star distribution of the Galactic bulge.  From
the Bulge$\_$GC1, M2FS CaT, and FLAMES CaT data, we found 59/82 (72$\%$),
191/466 (41$\%$), and 83/300 (28$\%$) stars to have velocities consistent
with cluster membership, respectively.  The significantly higher membership 
rate for the Bulge$\_$GC1 data is due to the preferential placement of fibers 
on stars closer to the cluster core.  Both CaT data sets also span a broader
color and luminosity range than the Bulge$\_$GC1 observations (see 
Figure \ref{f1}).

From the non--member distribution, we estimate that $\sim$0.5$\%$ of 
field stars will have a velocity between $+$120 and $+$170 km s$^{\rm -1}$ for 
the lines--of--sight probed here.  Since we have measured velocities for a 
total of 832 unique stars between the current data sets and Johnson et al. 
(2015a), we expect $\sim$5 field stars in the combined data to have velocities 
consistent with cluster membership.  However, the field star contamination 
rate may be overestimated because the cluster and field stars do not share the 
same spatial and metallicity distributions.  

Figures \ref{f1} and \ref{f3} show that a majority of stars having velocities 
consistent with cluster membership reside inside 4$\arcmin$ of the cluster 
center, but the obvious field stars are more uniformly distributed.  
Additionally, Johnson et al. (2015a) and Yong et al. (2016) have shown that 
most NGC 6273 stars have [Fe/H] $\la$ --1.35, but such stars are relatively 
rare in the bulge field (e.g., Zoccali et al. 2008; Bensby et al. 2013; Johnson
et al. 2013; Ness et al. 2013b).  The most likely contaminators are therefore 
stars that lie $\ga$ 4$\arcmin$ from the cluster center and have very red colors
and/or [Fe/H] $>$ --1.35.  Figure \ref{f3} indicates that 6 such stars exist 
in our data set.  Of these, stars 2MASS 17030978--2608035 and 17030625--2603576 
are the most likely to be field stars because both have [Fe/H] $>$ --0.8 and 
radial distances of $>$ 10$\arcmin$.  Star 2MASS 17024153--2621081 has 
[Fe/H] = --1.53, a radial distance of 5.1$\arcmin$, and is likely a cluster 
member.  The three remaining candidates (2MASS 17015056--2616256; 2MASS
17032450--2614557; 2MASS 17023960--2620224) have distances of 
4.3--10.6$\arcmin$ but lack [Fe/H] measurements so their membership cannot yet 
be confirmed.  Listings of star identifications, coordinates, photometry, and 
heliocentric radial velocities for member and non--member stars are provided 
in Tables 2 and 3, respectively.

Establishing membership near and beyond the tidal radius (14.57$\arcmin$; 
Alonso--Garc{\'{\i}}a et al. 2012) will be important in searches for any 
extended halo populations associated with NGC 6273, similar to what is observed
near clusters such as NGC 1851, M 2, NGC 5824, M 3, and M 13 (Grillmair et al.
1995; Olszewski et al. 2009; Marino et al. 2014b; Navin et al. 2015, 2016; 
Kuzma et al. 2016).  Figure \ref{f1} shows a possibly interesting morphology 
such that stars near the edge of our observations, which are also close to the 
tidal radius, are more numerous on the eastern side of the cluster than the 
western side.  However, more observations are needed to confirm that this 
asymmetry is real.

\subsection{Cluster Rotation}

\begin{figure}
\epsscale{1.00}
\plotone{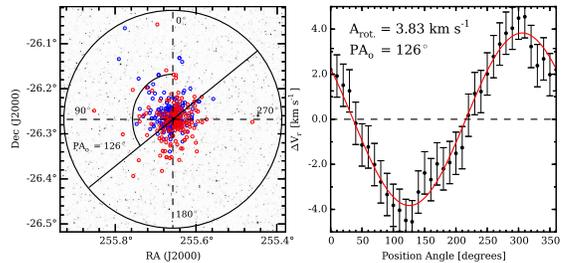}
\caption{\emph{Left:} the sky coordinates of member stars with heliocentric
radial velocities lower (blue) and higher (red) than the cluster average are
superimposed on a 2MASS J--band image.  The solid black line bisecting the
cluster illustrates the position angle of the rotation axis (PA$_{\rm o}$),
which is measured by rotating the solid black line east through west and
finding the maximum difference in heliocentric radial velocity on each side.
\emph{Right:} the average heliocentric radial velocity
difference for position angles measured in 10$\arcdeg$ increments.  The
solid red line indicates the best--fit sinusoidal function to the data.  See
text for details.}
\label{f4}
\end{figure}

Many globular clusters have been shown to rotate with amplitudes of order a 
few km s$^{\rm -1}$ (e.g., C{\^o}t{\'e} et al. 1995; Lane et al. 2009, 2010a;
Bellazzini et al. 2012; Bianchini et al. 2013; Kacharov et al. 2014; 
Kimmig et al. 2015; Lardo et al. 2015).  In Figure \ref{f4}, we investigated
net rotation in NGC 6273 by following a standard technique in which the average
radial velocity is calculated for stars on either side of an imaginary line
passing through the cluster center.  The bisecting line is rotated east through
west in 10$\degr$ increments, and the velocity differences are plotted as a 
function of position angle.  The resulting data can be fit with a sinusoidal
function of the form:

\begin{equation}
\Delta \langle V_{r} \rangle = A_{rot.}sin(PA + \Phi),
\end{equation}

\noindent 
where A$_{\rm rot.}$ is twice the actual projected rotation amplitude, 
$\Phi$ = 270$\degr$ -- PA$_{\rm o}$, and PA$_{\rm o}$ is the angle of maximum 
rotation.  Bellazzini et al. (2012) argue that the projected A$_{\rm rot.}$ 
value should be a reasonable estimate for the true maximum rotation amplitude, 
and we have adopted their interpretation here.

For NGC 6273, we find a clear rotation signature with 
A$_{\rm rot.}$ = 3.83 $\pm$ 0.12 km s$^{\rm -1}$ and PA$_{\rm o}$ = 126$\degr$
$\pm$ 2$\degr$.  We calculated the rotation profile using various angular bin 
sizes and found that while A$_{\rm rot.}$ only varied by a few tenths of a 
km s$^{\rm -1}$ the PA$_{\rm o}$ value could change by $\sim$15$\degr$.  
Therefore, we follow Bellazzini et al. (2012) and have adopted the conservative
1$\sigma$ uncertainties of $\pm$0.5 km s$^{\rm -1}$ for A$_{\rm rot.}$ and 
$\pm$30$\degr$ for PA$_{\rm o}$.  Compared to the large globular cluster 
samples presented in Bellazzini et al. (2012),  Kimmig et al. (2015), and 
Lardo et al. (2015), NGC 6273 exhibits relatively strong rotation.  NGC 6273's 
large A$_{\rm rot.}$ value is consistent with other clusters having similar 
metallicity and mass (e.g., $\omega$ Cen; see Figures 11 and 19 in Bellazzini 
et al. 2012 and Lardo et al. 2015, respectively).

\begin{figure}
\epsscale{1.00}
\plotone{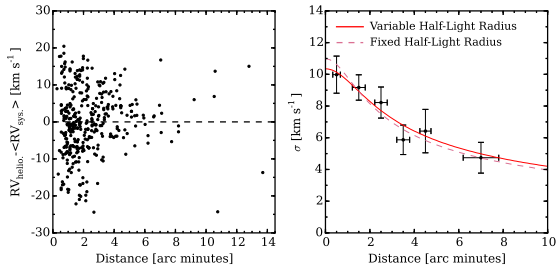}
\caption{\emph{Left:} the radial velocity difference between each star and the
cluster average is plotted as a function of the projected distance from the
cluster center.  \emph{Right:} the velocity dispersion for various radial bins
is plotted as a function of the projected distance from the cluster center.
Inside 5$\arcmin$, the data are binned into 1$\arcmin$ bins, and the last bin
includes all member stars with projected radial distances between
5--8.5$\arcmin$.  The outer bin is shown for context but was not included in
the fitting process.  The solid red line shows the best--fit Plummer model when
the central velocity dispersion and half--light radius are allowed to vary.
The dashed light red line shows the best--fit Plummer model when the
half--light radius is held fixed.  See text for details.}
\label{f5}
\end{figure}

In Figure \ref{f5}, we also investigated the change in velocity dispersion
as a function of the projected radial distance from the cluster center.  As
expected, we find that the velocity dispersion decreases from at least 
10 km s$^{\rm -1}$ inside 1$\arcmin$ to less than 5 km s$^{\rm -1}$ outside
5$\arcmin$.  We also estimated the cluster's central velocity dispersion 
($\sigma$$_{\rm o}$) using simple Plummer models (Plummer 1911) of the form:

\begin{equation}
\sigma^{2} (r) = \frac{\sigma _{o}^{2}}{\sqrt{1+(\frac{r}{r_{h}})^{2}}},
\end{equation}

\noindent
where r$_{\rm h}$ is the Plummer scale radius\footnote{As noted in
Lane et al. (2010b), the Plummer scale radius is equivalent to the projected
half--mass radius for projected Plummer models.}.  We fit two models: (1) one
with both $\sigma$$_{\rm o}$ and r$_{\rm h}$ varied as free parameters and (2)
one with $\sigma$$_{\rm o}$ varied as a free parameter and r$_{\rm h}$ held 
fixed.  For the latter case, we assumed the half--light radius was 
approximately equal to the half--mass radius and adopted a half--light radius 
of 1.32$\arcmin$ (Harris 1996; 2010 revision).  The resulting fit provided
$\sigma$$_{\rm o}$ = 10.98 $\pm$ 0.40 km s$^{\rm -1}$.  For the former case, 
we found $\sigma$$_{\rm o}$ = 10.35 $\pm$ 0.69 km s$^{\rm -1}$ and 
r$_{\rm h}$ = 1.67$\arcmin$ $\pm$ 0.41$\arcmin$.

However, we regard these values as lower limits of the true central velocity
dispersion because the measured velocity dispersion for the bin closest to
the cluster core is sensitive to the adopted bin size.  For example, when
the first bin contains stars with projected radial distances of 
0.2--1.0$\arcmin$, as is done in Figure \ref{f5}, the dispersion is $\sim$10
km s$^{\rm -1}$, but if we change the range to 0.2--0.7$\arcmin$ then the
dispersion increases to $\sim$12 km s$^{\rm -1}$.  Furthermore, a simple 
Plummer model assumes spherical symmetry, but NGC 6273 is relatively elliptical
in shape (White \& Shawl 1987; Chen \& Chen 2010).  Additional velocity
measurements inside $\sim$0.2--0.5$\arcmin$ and the application of more 
sophisticated models are likely to find a true $\sigma$$_{\rm o}$ $>$ 12
km s$^{\rm -1}$.  We estimate that the cluster's true 
A$_{\rm rot.}$/$\sigma$$_{\rm o}$ ratio is $\sim$0.30--0.35, which is typical
for massive elliptical metal--poor globular clusters (e.g., see Bellazzini et 
al. 2012; Kacharov et al. 2014; Kimmig et al. 2015; Lardo et al. 2015).

\section{SPECTROSCOPIC ANALYSIS}

\subsection{Model Atmospheres}

The model atmosphere parameters effective temperature (T$_{\rm eff}$), 
surface gravity (log(g)), metallicity ([Fe/H]), and microturbulence
($\xi$$_{\rm mic.}$) were determined spectroscopically for all radial velocity
member stars observed with the Bulge$\_$GC1 setup.  A spectroscopic 
determination of especially T$_{\rm eff}$ and log(g) is preferred over 
photometric measurements for NGC 6273 because of the cluster's large and 
variable reddening (see Section 2.3).  We followed the general analysis 
procedures outlined in Johnson et al. (2015a), which includes use of the 1D 
local thermodynamic equilibrium (LTE) line analysis code MOOG\footnote{The MOOG
source code is available at: http://www.as.utexas.edu/$\sim$chris/moog.html.} 
(Sneden 1973; 2014 version).  In particular, we solved for T$_{\rm eff}$ by 
enforcing excitation equilibrium with the \ion{Fe}{1} lines and solved for 
surface gravity by adjusting log(g) until the \ion{Fe}{1} and \ion{Fe}{2} lines 
provided the same abundance.  In the few instances where only \ion{Fe}{1} could
be measured, we assigned stars a log(g) value that was compatible with other 
cluster members of similar temperature and metallicity.  Microturbulence was 
measured by adjusting $\xi$$_{\rm mic.}$ until the derived 
log $\epsilon$(\ion{Fe}{1}) abundance was independent of line strength.  
Finally, the metallicity of each model was set as the average of 
[\ion{Fe}{1}/H] and [\ion{Fe}{2}/H].

In order to generate the models, we interpolated within the available grid
of ATLAS9 model atmospheres\footnote{The model atmosphere grid can be accessed
at: http://wwwuser.oats.inaf.it/castelli/grids.html.} (Castelli \& Kurucz 
2004).  For most stars, we used the $\alpha$--enhanced models in order to 
compensate for the difference between [Fe/H] and [M/H].  However, a small 
number of stars in our sample have [$\alpha$/Fe] $\sim$ 0, and for those stars
we used the scaled--solar models.  For every star, we started with a base--line
model of T$_{\rm eff}$ = 4500 K, log(g) = 1.20 cgs, [Fe/H] = --1.60 dex, and 
$\xi$$_{\rm mic.}$ = 1.70 km s$^{\rm -1}$, and iteratively solved for all four 
parameters simultaneously.

Lind et al. (2012) showed that for some stars departures from LTE can have
a significant impact on the model atmosphere parameters derived by 
spectroscopic methods.  However, the impact on stars in the temperature, 
gravity, and metallicity regime probed here is likely to be small.  
Additionally, the relative effects due to departures from LTE should be mostly 
negligible within a small parameter space (e.g., Wang et al. 2016), and we 
have attempted to empirically cancel out large non--LTE and 3D model atmosphere
deficiencies by performing a differential analysis relative to Arcturus.
Therefore, we have not applied any non--LTE corrections to our data.  We note
also that Dupree et al. (2016) showed the addition of a chromosphere may
alter the derived abundances for some elements.  However, since we lack the
spectral coverage necessary for constraining a chromospheric model, our 
model atmosphere parameters and abundances are based only on 
radiative/convective equilibrium models.  The final model atmosphere parameters
for all member stars derived from the Bulge$\_$GC1 data are provided in Table 4.

\subsection{Equivalent Width and Spectrum Synthesis Measurements}

The abundances of \ion{Si}{1}, \ion{Ca}{1}, \ion{Cr}{1}, \ion{Fe}{1}, 
\ion{Fe}{2}, and \ion{Ni}{1} were obtained by measuring the equivalent width 
(EW) of individual lines selected by Johnson et al. (2015a) to be relatively 
free of contamination from significant blends and residual telluric features.  
On average, the \ion{Si}{1}, \ion{Ca}{1}, \ion{Cr}{1}, \ion{Fe}{1}, 
\ion{Fe}{2}, and \ion{Ni}{1} abundances were based on the measurement of 2, 5, 
2, 33, 4, and 4 absorption lines, respectively.  However, we only measured
the abundances of these elements from the Bulge$\_$GC1 spectra.  We utilized 
the same EW measuring code, line list, and solar reference abundances 
described in Johnson et al. (2015a; see their Section 3.2 and their Table 2), 
and also used the same \emph{abfind} driver in MOOG to calculate the final 
abundance ratios. The [\ion{Si}{1}/Fe], [\ion{Ca}{1}/Fe], [\ion{Cr}{1}/Fe], 
[\ion{Fe}{1}/H], [\ion{Fe}{2}/H], and [\ion{Ni}{1}/Fe] abundances for every 
cluster member observed in the Bulge$\_$GC1 setup are provided in Tables 5--6.

The abundances of \ion{Na}{1}, \ion{Mg}{1}, \ion{Al}{1}, \ion{La}{2}, and 
\ion{Eu}{2} were obtained by using the \emph{synth} driver in MOOG to fit 
synthetic spectra to the observations.  The synthetic spectra were calculated 
using the line list developed for Johnson et al. (2015a), which is tuned to 
reproduce the Arcturus spectrum near the lines of interest and includes the 
updated CN line list from Sneden et al. (2014).  We preferred to 
use spectrum synthesis rather than an EW analysis for these elements because 
their abundances are more sensitive to blending, contamination from other 
features, and/or broadening effects.  For example, the Na and Al lines can have
significant contamination from nearby atomic features and molecular CN, 
especially in the more metal--rich stars.  Additionally, the Mg triplet near 
6319 \AA\ contains very weak lines, and the nearby continuum can be affected by
a shallow but broad \ion{Ca}{1} autoionization feature.  The La and Eu lines 
are also relatively weak, but are further affected by hyperfine structure 
broadening.  The Eu lines also contain a mixture of transitions from the 
\iso{151}{Eu} and \iso{153}{Eu} isotopes, for which we assumed the 
\iso{151}{Eu}:\iso{153}{Eu} Solar System ratio of 47.8$\%$:52.2$\%$ (Lawler et 
al. 2001).  

The final [Na/Fe], [Mg/Fe], [Al/Fe], [La/Fe], and [Eu/Fe] abundances derived 
for cluster members observed with the Bulge$\_$GC1 setup are provided in 
Tables 5--6.  All atomic parameters and solar reference abundances are 
available in Johnson et al. (2015a; their Table 2).

\subsection{Calcium Triplet Abundances}

In addition to the [Fe/H] abundances derived from the EW measurements of 
individual \ion{Fe}{1} and \ion{Fe}{2} lines, we measured [Fe/H] in a larger 
sample of stars using the 8542 \AA\ and 8662 \AA\ CaT lines.  These strong 
lines have been shown to be sensitive to a star's metallicity and relatively 
insensitive to a star's age or [$\alpha$/Fe] abundance, in a variety of 
environments (e.g., Armandroff \& Da Costa 1991; Olszewski et al. 1991; Idiart 
et al. 1997; Rutledge et al. 1997; Cole et al. 2004; Carrera et al. 2007; 
Battaglia et al. 2008; Da Costa 2016b).  Although several CaT--metallicity 
calibrations exist (e.g., Starkenburg et al. 2010; Saviane et al. 2012; Carrera
et al. 2013; V{\'a}squez et al. 2015), we followed the technique outlined in 
Yong et al. (2016) that utilizes the Mauro et al. (2014) calibration.  

As noted by Yong et al. (2016), the Mauro et al. (2014) calibration has two
significant advantages for NGC 6273: (1) the luminosity component of the 
calibration depends on a star's K$_{\rm S}$ magnitude, rather than V magnitude,
which is much less affected by differential reddening; and (2) the 
significantly flatter slope of the summed EW ($\Sigma$EW) versus 
K$_{\rm S}$(HB)--K$_{\rm S}$ relation reduces the effects of photometric,
distance, and reddening uncertainties on the derived [Fe/H] values.  
Additionally, we note that 2MASS provides uniform K$_{\rm S}$ photometry for 
our entire sample, but uniform V magnitudes are not yet available for all
stars.  However, since most CaT--metallicity relations may only be reliable 
down to the luminosity level of the HB (e.g., Da Costa et al. 2009), we did not
determine CaT metallicities for stars fainter than the HB.  This cut--off
primarily affected the FLAMES CaT sample.

The Mauro et al. (2014) calibration requires a measurement of the summed 
8542 \AA\ and 8662 \AA\ CaT EWs, defined as:

\begin{equation}
\Sigma EW = EW_{8542} + EW_{8662},
\end{equation}

\noindent
and the value K$_{\rm S}$(HB)--K$_{\rm S}$, where K$_{\rm S}$(HB) is the 
magnitude of the horizontal branch.  Following Yong et al. (2016), we have 
adopted K$_{\rm S}$(HB) = 12.85 magnitudes (Valenti et al. 2007).  The EWs for 
each line were fit using a function that is the sum, rather than the 
convolution, of a Gaussian and Lorentzian profile.  Using Equation 7 and 
following Mauro et al. (2014), we adopted their relation,

\begin{equation}
\Sigma EW = -0.385[K_{S}(HB) - K_{S}] + W^{\prime},
\end{equation}

\noindent
to solve for the reduced equivalent width (W$^{\prime}$).  The [Fe/H] values
for each star were then determined using Equation 8 and the cubic calibration 
from Mauro et al. (2014) for the Carretta et al. (2009a) metallicity scale:

\begin{equation}
[Fe/H] = -4.61 + 1.842\langle {W}'\rangle - 0.4428\langle {W}'\rangle^{2} + 0.04517\langle {W}'\rangle^{3}.
\end{equation}

\noindent
The individual EWs, $\Sigma$EW, and W$^{\prime}$ values for all NGC 6273 
members are provided in Table 7.

\begin{figure}
\epsscale{1.00}
\plotone{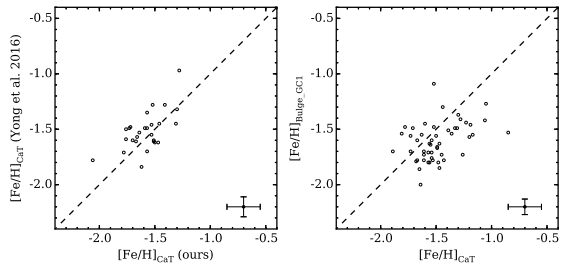}
\caption{\emph{Left:} a comparison of the CaT [Fe/H] values derived in this
work and Yong et al. (2016), for 27 stars in common.  The dashed line indicates
perfect agreement.  \emph{Right:} a comparison of the [Fe/H] values derived
from the CaT and Bulge$\_$GC1 data sets of this work, for 50 stars in common.
Note that in both panels our CaT [Fe/H] values are those derived from
Equation 9 and have not yet been corrected to place the CaT [Fe/H] abundances
on the Bulge$\_$GC1 [Fe/H] scale.  Typical error bars are shown in the bottom
right corner of each panel.}
\label{f6}
\end{figure}

A comparison between the observations of Yong et al. (2016) and our CaT data
set revealed 27 stars in common.  For this subset, the Yong et al. (2016)
[Fe/H] values are on average 0.06 dex more metal--rich than ours, but the 
metallicities from both studies are well--correlated (see Figure \ref{f6}).
Similarly, we found 50 stars in our sample that were observed in both the 
CaT and Bulge$\_$GC1 setups, and a comparison of the derived [Fe/H] values is
provided in Figure \ref{f6}.  The [Fe/H] measurements from both data sets
are relatively well--correlated, but the CaT data are 0.12 dex more 
metal--rich, on average.  Therefore, the final CaT--based [Fe/H] abundances 
provided in Table 7, and used throughout the rest of the paper, have been 
shifted by --0.12 dex in order to place the CaT and Bulge$\_$GC1 data sets on 
the same scale. 

\subsection{Internal Abundance Uncertainties}

For the reasonably high S/N Bulge$\_$GC1 region spectra analyzed here, the 
dominant sources of internal abundance uncertainties are related to the 
line--to--line abundance scatter from uncertain log(gf) values, small profile
fitting and/or continuum placement errors, and model atmosphere parameter
uncertainties.  The standard error of the mean provides a reasonable estimate
of the abundance errors due to line list and profile fitting uncertainties, 
and for this data set we find a typical measurement uncertainty of 0.05 dex 
($\sigma$ = 0.03 dex) in log $\epsilon$(X).  

\begin{figure}
\epsscale{1.00}
\plotone{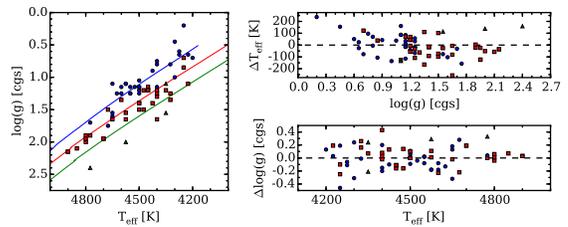}
\caption{The left panel shows the T$_{\rm eff}$ and log(g) values derived here
and in Johnson et al. (2015) for the metal--poor ([Fe/H] $<$ --1.65; filled
blue circles), metal--intermediate (--1.65 $<$ [Fe/H] $\leq$ --1.35; filled
red boxes), and metal--rich ([Fe/H] $>$ --1.35; filled green triangles)
populations, and compares the spectroscopic parameters with those predicted by
Dartmouth isochrones (Dotter et al. 2008).  The isochrones have an age of 12
Gyr, [$\alpha$/Fe] = 0.4 dex, and [Fe/H] = --1.75 (blue line), --1.50 (red
line), and --1.20 dex (green line).  The top right panel compares the
differences between the spectroscopic and isochrone temperatures
($\Delta$T$_{\rm eff}$) for a given surface gravity.  Similarly, the bottom
right panel compares the differences between the spectroscopic and isochrone
surface gravities ($\Delta$log(g)) for a given temperature.}
\label{f7}
\end{figure}

In order to estimate the uncertainties in T$_{\rm eff}$ and log(g), we provide
a comparison of the spectroscopically derived parameters with those expected
from Dartmouth isochrones (Dotter et al. 2008) with ages of 12 Gyr, 
[$\alpha$/Fe] = $+$0.4 dex, and [Fe/H] = --1.75, --1.50, and --1.20 dex in
Figure \ref{f7}.  The isochrones with different [Fe/H] are included because of
the metallicity spread detected in the cluster (Johnson et al. 2015a; Han et 
al. 2015; Yong et al. 2016; see also Section 5.1).  Figure \ref{f7} shows that
the derived temperature and surface gravity values are in good agreement with
those predicted by the isochrones.  Specifically, we find the average 
differences between the spectroscopic and isochrone temperature 
($\Delta$T$_{\rm eff}$) and surface gravity ($\Delta$log(g)) values to be 
--8 K and $+$0.01 cgs, respectively, and do not detect any significant trends 
as a function of temperature, gravity, or metallicity.  The dispersions in
$\Delta$T$_{\rm eff}$ and $\Delta$log(g) are found to be 92 K and 0.17 cgs,
respectively.  Therefore, we have adopted 100 K and 0.15 cgs as the typical
model atmosphere uncertainties for T$_{\rm eff}$ and log(g).  For the 
model atmosphere metallicity, we have adopted an uncertainty of 0.10 dex based
on the combined measurement errors of [\ion{Fe}{1}/H] and [\ion{Fe}{2}/H].
Additionally, we estimate the typical $\xi$$_{\rm mic.}$ uncertainty to be
0.10 km s$^{\rm -1}$ based on the scatter and fitting uncertainties present in 
plots of log $\epsilon$(\ion{Fe}{1}) versus log(EW/$\lambda$).

The abundance uncertainty values ($\Delta$[X/Fe] or $\Delta$[Fe/H]) were 
determined by rerunning MOOG and changing each model atmosphere parameter by 
the estimated uncertainties listed previously.  Only one parameter was changed 
per run while the other values were held fixed.  To speed up the analysis, we 
converted abundances to EWs for the elements measured by spectrum synthesis
using the \emph{ewfind} driver in MOOG.  The total internal abundance 
uncertainties listed in Tables 5--6 were determined by adding the model 
atmosphere error terms, plus the random measurement uncertainties, in 
quadrature.  

For the CaT data, we estimated the abundance uncertainties by analyzing the 
correlation between the 8542 \AA\ and 8662 \AA\ EWs.  In other words, we 
used the strong correlation between EW$_{\rm 8542}$ and EW$_{\rm 8662}$ to
predict the EW of each line based on the other one.  The predicted EWs were 
then propagated through Equations 8--9, and the difference between these values
and the [Fe/H] abundance given in Table 7 was taken as the measurement error.  
Using this method, we found an average $\Delta$[Fe/H] = 0.15 dex ($\sigma$ = 
0.07 dex).  We note that this value is similar to the fitting uncertainty of 
Equation 9 (Mauro et al. 2014).  A typical [Fe/H] uncertainty of 0.15 dex is 
also similar to the 1$\sigma$ scatter (0.21 dex) between [Fe/H] values 
determined from the CaT and Bulge$\_$GC1 data.  

\section{RESULTS AND DISCUSSION}

\subsection{Metallicity Distribution}

The color and CaT abundance spreads observed by Piotto et al. (1999) and 
Rutledge et al. (1997) provided some of the first evidence that NGC 6273 may
host stars with different metallicities.  More recently, high resolution 
spectroscopic measurements from Johnson et al. (2015a) showed that the cluster
contains stars with [Fe/H] ranging from --1.80 to --1.30 dex, and also found 
that the cluster hosts at least two distinct populations separated in [Fe/H] by
$\sim$0.25 dex.  Similarly, Han et al. (2015) used narrow--band \emph{hk}
photometry to clearly show that the cluster's sub--giant branch and RGB are 
split into two sequences with different compositions.  Yong et al. (2016) also 
reported CaT metallicities ranging from [Fe/H] = --1.84 to --0.70 dex, further
indicating the presence of a large metallicity spread in the cluster.

\begin{figure}
\epsscale{1.00}
\plotone{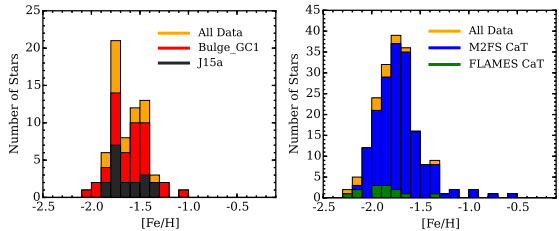}
\caption{The left and right panels compare the [Fe/H] distribution functions
derived from data obtained with the Bulge$\_$GC1 and CaT spectrograph setups,
respectively.  For the left panel, the orange histogram represents the sum of
the metallicities derived from this work and Johnson et al. (2015a).
Similarly, in the right panel the orange histogram represents the sum of the
M2FS and FLAMES CaT metallicities.  All of the data are sampled with 0.10 dex
[Fe/H] bins.  Note the broad [Fe/H] range found in both data sets, and also
the likely presence more than one distinct population in the Bulge$\_$GC1 data
set.  The [Fe/H] distributions in both panels only include stars that are
radial velocity members.}
\label{f8}
\end{figure}

The Johnson et al. (2015a) and Yong et al. (2016) spectroscopic results are 
based on the analysis of only 18 and 44 RGB stars observed in the Bulge$\_$GC1 
and CaT spectral regions, respectively.  Therefore, we add here 
[Fe/H] measurements for 51 RGB members in the Bulge$\_$GC1 region and 191 RGB 
members in the CaT region (see Tables 4, 6, and 7).  For the Bulge$\_$GC1 
data, we find a full range of [Fe/H] = --2.00 to --1.09 dex, an average 
$\langle$[Fe/H]$\rangle$ = --1.61 dex, a dispersion ($\sigma$$_{\rm [Fe/H]}$) 
of 0.18 dex, and an interquartile range (IQR) of 0.24 dex.  Similarly, the CaT 
data exhibit a full range of [Fe/H] = --2.22 to --0.56 dex, an average 
$\langle$[Fe/H]$\rangle$ = --1.73 dex, $\sigma$$_{\rm [Fe/H]}$ = 0.24 dex, and 
an IQR of 0.27 dex.  A comparison between the Bulge$\_$GC1 and CaT metallicity
distributions is shown in Figure \ref{f8}, and both data sets provide evidence 
that NGC 6273 harbors an intrinsic metallicity spread.

Further examination of Figure \ref{f8} also indicates that NGC 6273 may host
distinct populations with different [Fe/H], rather than just a broadened 
distribution.  Specifically, Figure \ref{f8} suggests that at least three
major components may exist: (1) a ``metal--poor" population ([Fe/H] $\leq$ 
--1.65); (2) a ``metal--intermediate" population (--1.65 $<$ [Fe/H] $\leq$ 
--1.35); and a ``metal--rich" tail ([Fe/H] $>$ --1.35), and that these 
components constitute 46$\%$ $\pm$ 8$\%$, 48$\%$ $\pm$ 8$\%$, 6$\%$ $\pm$ 4$\%$
of our total Bulge$\_$GC1 data set, respectively.  We find the average 
metallicities of the metal--poor, metal--intermediate, and metal--rich 
populations to be: $\langle$[Fe/H]$\rangle$ = --1.77 dex ($\sigma$ = 0.08 dex),
$\langle$[Fe/H]$\rangle$ = --1.51 dex ($\sigma$ = 0.07 dex), and 
$\langle$[Fe/H]$\rangle$ = --1.22 dex ($\sigma$ = 0.09 dex), respectively.
The clustering of stars near [Fe/H] = --1.75 and --1.50 is consistent
with the [Fe/H] abundances and split RGB sequences derived by Johnson et al.
(2015a) and Han et al. (2015), and the presence of a metal--rich tail extending
up to at least [Fe/H] $\approx$ --1 matches the findings of Yong et al. (2016).

\begin{figure}
\epsscale{1.00}
\plotone{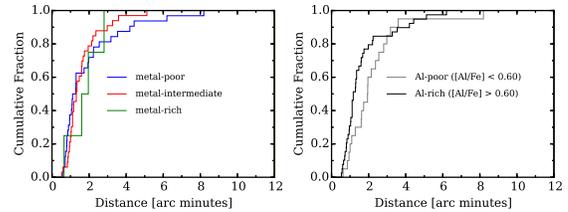}
\caption{\emph{Left:} the cumulative distribution functions of the metal--poor
(blue), metal--intermediate (red), and metal--rich (green) populations are
shown as a function of the projected distance from the cluster center.
Note the preferential central concentration of metal--intermediate, and
possibly metal--rich, stars at distances $\ga$ 1.5$\arcmin$.  \emph{Right:} a
similar plot comparing the radial distributions of Al--poor (grey; ``first
generation") and Al--rich (black; ``second generation") stars from all three
major populations.  Note the significant central concentration of Al--rich
stars.}
\label{f9}
\end{figure}

Figure \ref{f9} indicates that a radial metallicity gradient may exist
in the cluster such that the metal--intermediate stars are more centrally
concentrated than the metal--poor stars.  Although the metal--rich stars 
observed with the Bulge$\_$GC1 setup all reside inside 3$\arcmin$ of the 
cluster center (see also Figure \ref{f3}), the sample size is too small to 
draw any strong conclusions about this population's radial distribution.  For
the two dominate populations, a difference in their radial distributions is
only  observed at projected distances $\ga$ 1.5$\arcmin$ from the cluster 
center, and a two--sided Kolmogorov--Smirnov test indicates that we do not have
enough evidence to reject the null hypothesis that the two data sets are drawn 
from the same radial distribution.  However, we note that the radial range where
the distributions may differ is within $\sim$1--3 half--mass radii, which is the
region that Vesperini et al. (2013) estimate the local population mixtures
may closely match the global ratios.  Interestingly, if the radial segregation 
of stars with different metallicities is confirmed from larger sample sizes, 
then NGC 6273 would share a similar metallicity gradient morphology with 
$\omega$ Cen (e.g., Norris et al. 1996; Suntzeff \& Kraft 1996; Rey et al. 
2004; Bellini et al. 2009; Johnson \& Pilachowski 2010).  Such a gradient would
contrast with NGC 1851 where Carretta et al. (2010b) found the metal--poor 
stars to be the most centrally concentrated.

\subsection{Additional Evidence of a Complex Metallicity Distribution}

Spectroscopic observations have indicated that several clusters, including
$\omega$ Cen (e.g., Norris \& Da Costa 1995; Johnson \& Pilachowski 2010; 
Marino et al. 2011a), NGC 5286 (Marino et al. 2015), M 2 (Yong et al. 2014), 
M 54 (Carretta et al. 2010a), Terzan 5 (Origlia et al. 2013; Massari et al. 
2014), NGC 1851 (Yong \& Grundahl 2008; Carretta et al. 2011; Lim et al. 
2015), and M 22 (Pilachowski et al. 1982; Da Costa et al. 2009; Marino et al. 
2009, 2011b), may host multiple populations with distinct [Fe/H] ratios.  
However, recent studies by Mucciarelli et al. (2015a) and Lardo et al. (2016) 
claim that at least some of these [Fe/H] spreads are spurious detections driven
by a disparity between [\ion{Fe}{1}/H] and [\ion{Fe}{2}/H].  Similarly, Ivans 
et al. (2001), Lapenna et al. (2014), and Mucciarelli et al. (2015b) found that
[Fe/H] determinations for RGB and AGB stars can differ systematically by 
$>$ 0.1 dex, and that mixing RGB and AGB stars in a sample can produce an 
artificial metallicity spread.  A common thread connecting these issues is the 
method by which a star's surface gravity is determined (spectroscopic versus 
photometric).  Specifically, spectroscopic determinations may produce optimal 
log(g) values that correspond to masses which are systematically too low 
($<$ 0.5 M$_{\odot}$ in many cases).  Since we utilize a spectroscopic 
surface gravity method and find that NGC 6273 shares many chemical and 
morphological characteristics with clusters such as M 2 and M 22, for which
intrinsic [Fe/H] spreads are contested, it is prudent to examine alternative
lines of evidence that may support or refute NGC 6273 possessing an intrinsic
metallicity spread.

\begin{figure}
\epsscale{1.00}
\plotone{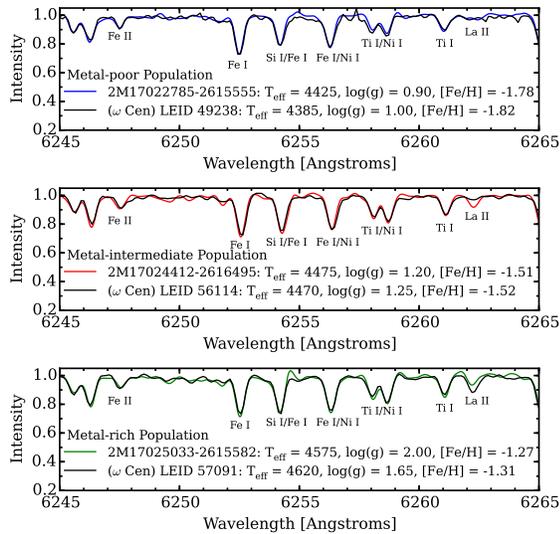}
\caption{This figure compares the spectra of stars in NGC 6273 and $\omega$ Cen
(Johnson \& Pilachowski 2010) that have similar T$_{\rm eff}$, log(g), and
[Fe/H].  The top, middle, and bottom panels show stars from the metal--poor,
metal--intermediate, and metal--rich groups, respectively.  In these panels,
the colored spectra are from stars in NGC 6273 and the black spectra are from
stars in $\omega$ Cen.  The NGC 6273 M2FS spectra have been smoothed to match
the resolution of the $\omega$ Cen Hydra spectra (R $\sim$ 18,000).}
\label{f10}
\end{figure}

Both spectroscopy and photometry unambiguously agree that $\omega$ Cen 
possesses discrete RGB populations with different [Fe/H], and in 
Figure \ref{f10} we directly compare the spectra of NGC 6273 and $\omega$ Cen 
stars that have physical parameters typical of those in the metal--poor, 
metal--intermediate, and metal--rich groups.  The $\omega$ Cen temperature and 
gravity parameters were determined entirely from photometric methods (see 
Johnson \& Pilachowski 2010), assuming masses of 0.8 M$_{\odot}$, and 
therefore should avoid the potential spectroscopic gravity problems noted 
above.  As can be seen in Figure \ref{f10}, the nearly identical \ion{Fe}{1} 
and \ion{Fe}{2} line profiles suggest that the NGC 6273 and $\omega$ Cen stars 
share similar compositions across a wide [Fe/H] range.  We note also that the 
CaT [Fe/H] distribution shown in Figure \ref{f8} for NGC 6273 closely matches 
the extended metallicity distribution found in $\omega$ Cen (e.g., Norris et 
al. 1996; Suntzeff \& Kraft 1996).

\begin{figure}
\epsscale{1.00}
\plotone{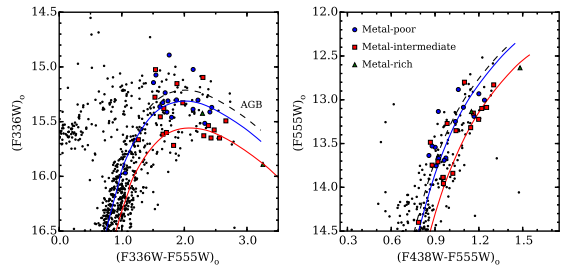}
\caption{The left and right panels compare the upper RGB and AGB regions of
NGC 6273 using combinations of the dereddened F336W, F438W, and F555W bands.
The Bulge$\_$GC1 observations that overlap with the WFC3 field--of--view are
distinguished by metallicity using the same criteria, colors, and symbols as
those in Figure \ref{f7}.  The two panels include Dartmouth isochrones with
ages of 12 Gyr, [$\alpha$/Fe] = 0.4 dex, distances of 9 kpc, and [Fe/H] =
--1.75 (blue lines) and --1.50 (red lines) dex, which correspond to the
metallicities of the two dominant populations.  The dashed black lines separate
the RGB and AGB stars.  The bluer colors of the most metal--rich stars (green
triangles) suggest that these stars may have different He, C, N, and O
abundances than the metal--intermediate population and/or may be higher
metallicity AGB stars.}
\label{f11}
\end{figure}

As shown in Figure \ref{f2}, the broad color dispersion along the upper RGB
provides some evidence that NGC 6273 may harbor an intrinsic metallicity 
spread.  We investigate this further in Figure \ref{f11} by examining the upper
RGB regions of the (F336W)$_{\rm o}$ versus (F336W--F555W)$_{\rm o}$ and 
(F555W)$_{\rm o}$ versus (F438W--F555W)$_{\rm o}$ color--magnitude diagrams and
identifying the Bulge$\_$GC1 spectroscopic targets with different [Fe/H].  Both
color--magnitude diagrams indicate that the two dominant metallicity groups 
tend to separate on the upper RGB.  The (F336W)$_{\rm o}$ versus 
(F336W--F555W)$_{\rm o}$ plot in particular suggests that the brightest $\sim$ 
0.5 magnitudes of the RGB--tip may split into at least two sequences with 
different [Fe/H], which is similar to the result found by Marino et al. (2015) 
for NGC 5286.  However, the F336W and F438W filters, and by extension the 
F336W--F555W and F438W--F555W colors, can be sensitive to both a star's overall
metallicity and its C$+$N$+$O abundances.  Therefore, the color--magnitude 
diagrams shown in Figure \ref{f11} are consistent with an intrinsic metallicity
spread, but a detailed examination of the cluster's CNO (and also He) 
abundances is required in order to fully confirm this result.  Interestingly, 
the two metal--rich stars in Figure \ref{f11} are located at colors that are 
bluer than might be expected from their metallicities alone.  We note that 
similar observations have been found for the equivalent ``s--poor/Fe--rich"
stars in NGC 5286 (Marino et al. 2015) and M 2 (Yong et al. 2014).  The bluer
colors for these stars may reflect lower atmospheric opacities driven by
different light element compositions and perhaps lower [$\alpha$/Fe] ratios, 
at least for NGC 6273\footnote{We note that CNO variations are likely present
in NGC 6273 since Han et al. (2015) found the more Ca--rich (metal--rich) stars
to have enhanced CN and CH.  Additionally, a few of the most metal--rich stars 
in our data set have very strong CN lines.}.  

Figure \ref{f11} also shows that very few of our Bulge$\_$GC1 targets are on 
the AGB, indicating that the measured metallicity spread is not caused by 
systematic differences in the RGB and AGB abundance scales.  In fact, the few 
AGB stars in our sample appear to belong to both the metal--poor and 
metal--intermediate populations, and the two metal--rich stars could also
belong to a more metal--rich AGB sequence.  Therefore, we regard the combined 
evidence of separate subgiant and RGB sequences observed by Han et al. (2015)
with the \emph{hk} filter, the RGB color dispersions seen in Figures \ref{f2} 
and \ref{f11} here, the large metallicity spreads detected previously by Johnson
et al. (2015a) and Yong et al. (2016), and the spectroscopic data presented 
here as strong evidence that NGC 6273 possesses an intrinsic metallicity spread.

\subsection{Light and Heavy Element Chemical Abundance Patterns}

\subsubsection{Alpha Element Abundances}

The $\alpha$ elements Mg, Si, and Ca are largely produced during hydrostatic
and explosive carbon, neon, and oxygen burning in massive stars (e.g., Woosley
\& Weaver 1995).  In environments where chemical enrichment has been dominated
by the products of core--collapse supernovae (SNe), one tends to find stars 
with [$\alpha$/Fe] abundances that are enhanced by about a factor of 2--3 over 
the solar ratio (e.g., see review by McWilliam 1997).  In contrast, longer 
enrichment time scales may produce stars with lower [$\alpha$/Fe] ratios as
Type Ia SNe begin to contribute larger amounts of Fe--peak elements than 
$\alpha$ elements (Tinsley 1979). 

As can be clearly seen in Gratton et al. (2004; their Figure 4), nearly all 
Galactic globular clusters have [$\alpha$/Fe] $\sim$ 0.2--0.4 dex.  
Furthermore, the star--to--star scatter of [$\alpha$/Fe] within a given cluster
is typically $<$ 0.1 dex, which suggests that the products of core--collapse
SNe were well--mixed.  Only a small number of clusters, such as Ruprecht 106,
Terzan 7, and Palomar 12, are known to have abnormally low 
[$\alpha$/Fe] ratios (e.g., Pritzl et al. 2005), and all three of these 
clusters are thought to have extragalactic/accretion origins (e.g., Cohen 2004;
Law \& Majewski 2010; Villanova et al. 2013).  Therefore, one does not normally
expect to find stars with enhanced and depleted [$\alpha$/Fe] ratios within a 
single globular cluster, beyond the well--known proton--capture nucleosynthesis
variations (see Section 5.3.2).  In fact, only the massive iron--complex 
clusters $\omega$ Cen, NGC 6273, M 54, M 2, and Terzan 5 show any evidence of 
hosting stars with different [$\alpha$/Fe] ratios (Pancino et al. 2002; Origlia
et al. 2003, 2011, 2013; Carretta et al. 2010a; Johnson \& Pilachowski 2010; 
Yong et al. 2014; Johnson et al. 2015a).

\begin{figure}
\epsscale{1.00}
\plotone{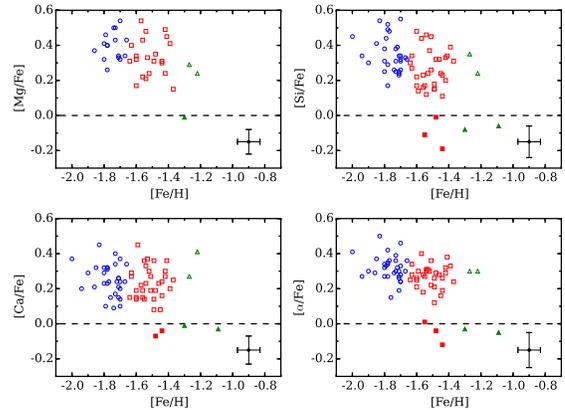}
\caption{The [Mg/Fe], [Si/Fe], [Ca/Fe], and [$\alpha$/Fe] ratios for NGC 6273
stars observed in this work and Johnson et al. (2015a) are plotted as a
function of [Fe/H].  The open blue circles, red boxes, and green triangles
designate stars belonging to the metal--poor, metal--intermediate,  and
metal--rich populations, respectively.  The filled symbols indicate stars that
have low [Mg/Fe], [Si/Fe], [Ca/Fe], and [$\alpha$/Fe] abundances.  The dashed
black lines indicate the solar [X/Fe] ratios, and the [$\alpha$/Fe] abundances
represent the average values of [Mg/Fe], [Si/Fe], and [Ca/Fe] measured in each
star.  Typical error bars are included in the bottom right corner of each plot.}
\label{f12}
\end{figure}

Figure \ref{f12} and Table 8 show the [Mg/Fe], [Si/Fe], [Ca/Fe], and averaged
[$\alpha$/Fe] patterns of NGC 6273's various populations.  In agreement with
Johnson et al. (2015a), we find that most stars in NGC 6273 have elevated 
$\alpha$ element abundances, but that the average [Mg/Fe] and [Si/Fe] 
ratios may decrease slightly as a function of increasing metallicity.  The
[Si/Fe] abundances in particular may show additional substructure, and we find
some evidence that the average [Si/Fe] abundances of the ``$\alpha$--enhanced" 
metal--intermediate stars may be lower than those of the metal--poor and 
metal--rich groups.  We note that a similar change in the [Si/Fe] abundances
with [Fe/H] has been observed in $\omega$ Cen (Johnson \& Pilachowski 2010),
which suggests that this trend could be the signature of a particular 
self--enrichment mode in massive clusters.  However, the [Ca/Fe] abundances 
show no significant trends as a function of [Fe/H], and the typical dispersion 
within each sub--population is $\sim$0.1 dex.  

In a previous analysis, Johnson et al. (2015a) discovered that the most 
metal--rich star in their sample exhibited low [X/Fe] ratios for several 
species, including the $\alpha$ elements.  The new data presented here indicate
that not all metal--rich stars have low [$\alpha$/Fe], but we find at least 
five ``low--$\alpha$" stars that have approximately solar [Mg/Fe], [Si/Fe], 
and [Ca/Fe] abundances.  As can be seen in Figure \ref{f12}, all five 
low--$\alpha$ stars have [Fe/H] $>$ --1.5 dex.  Additionally, the specific 
frequency of low--$\alpha$ stars increases with metallicity such that these 
stars constitute 9$\%$ (3/33) of the metal--intermediate population and 50$\%$ 
(2/4) of the metal--rich population.  However, we caution that the measured 
specific frequency values are likely affected by small number statistics and 
should be confirmed with additional observations.

Although we noted above that $\omega$ Cen, M 54, M 2, and Terzan 5 also contain 
stars with higher [Fe/H] and lower [$\alpha$/Fe], none of these clusters 
exactly matches the pattern of NGC 6273.  For example, the $\alpha$--poor 
stars in M 2 and Terzan 5 are exclusively found in the most metal--rich 
populations, but neither cluster appears to contain $\alpha$--enhanced 
\emph{and} $\alpha$--poor stars at the same metallicity.  Although Johnson \& 
Pilachowski (2010; see their Figure 10) found several stars with high and 
low [Si/Fe] and [Ca/Fe] abundances across a broad range of [Fe/H] in 
$\omega$ Cen, follow--up observations are required to confirm that this pattern
matches what is found in NGC 6273.  

Interestingly, the M 54 cluster and Sagittarius nuclear field star system may 
provide the closest example to what is observed in NGC 6273 (see Carretta et 
al. 2010a).  In this system, the metal--poor cluster M 54 contains a metallicity
spread but only $\alpha$--enhanced stars.  In contrast, the surrounding galaxy 
field stars are generally more metal--rich and have lower [$\alpha$/Fe].  
Therefore, if NGC 6273 formed in the core of a system similar to the 
Sagittarius dwarf galaxy, then the cluster may have been able to accrete
a small number of metal--rich, $\alpha$--poor field stars from its progenitor 
population.  Alternatively, the low--$\alpha$ stars in NGC 6273 may have been 
preferentially polluted by the ejecta of Type Ia SNe, perhaps in a scenario 
similar to that discussed in D'Antona et al. (2016).  However, such a scenario 
would have to be able to produce low--$\alpha$ stars with different [Fe/H] but 
otherwise similar compositions (see Sections 5.3.2 and 5.3.3), and may even 
have to occur multiple times in clusters like NGC 6273.

\subsubsection{Light Element Abundances}

As mentioned in Section 1, globular clusters show clear light element abundance
variations that extend beyond the effects of first dredge--up and are a result
of high temperature ($>$ 40 MK; Langer et al. 1993, 1997; Prantzos et al. 2007)
proton--capture burning.  Since these effects are observed in main--sequence 
and evolved RGB stars (e.g., Gratton et al. 2001), we know that the gas from 
which present day cluster stars formed was polluted by a previous generation of
more massive stars.  Although the exact nucleosynthesis sources remain a 
mystery, the observed effects include anti--correlations among the element 
pairs C--N, O--N, O--Na, O--Al, and Mg--Al and correlations of C--O, N--Na, 
and Na--Al (e.g., Sneden et al. 2004).  He enhancements are also likely found 
in stars with low O/Mg and high Na/Al (e.g., Bragaglia 2010a,b; Dupree et al. 
2011).  For this paper, we adopt the common nomenclature that ``first 
generation" stars are those with compositions similar to metal--poor halo field
stars (i.e., lower He, N, Na, and Al abundances; higher C, O, and Mg 
abundances) and ``second generation" stars are those with enhanced He, N, Na, 
and Al abundances and depleted C, O, and possibly Mg abundances.  

The Mg--Al anti--correlation is only found in a handful of the most massive 
clusters, but may be particularly useful for identifying discrete populations 
(e.g., Carretta 2014, 2015).  Since the full Mg--Al cycle is activated at a 
higher temperature than the O--N and Ne--Na cycles, the presence (or not) of a 
Mg--Al anti--correlation provides important insight into the burning 
temperatures achieved by the pollution sources.  Similarly, a few of the most 
massive clusters also exhibit abundance variations that extend to elements as 
heavy as Si, K, and Sc, which is likely a byproduct of even higher temperature 
proton--capture burning (Yong et al. 2005; Carretta et al. 2009b, 2013, 2014; 
Johnson \& Pilachowski 2010; Cohen \& Kirby 2012; Mucciarelli et al. 2012, 
2015c; Ventura et al. 2012; Carretta 2015; Roederer \& Thompson 2015).  
Notably, many of these clusters share similar properties with NGC 6273, such 
as extended blue HBs.

\begin{figure}
\epsscale{1.00}
\plotone{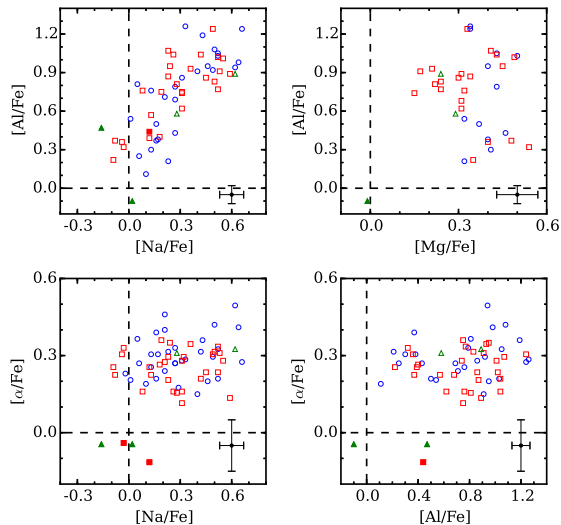}
\caption{\emph{Top:} these panels show the NGC 6273 [Al/Fe] abundances from
this work and Johnson et al. (2015a) plotted as a function of [Na/Fe] (left)
and [Mg/Fe] (right).  A clear Na--Al correlation exists for all three
metallicity groups, but a Mg--Al anti--correlation may only be present in the
metal--intermediate and metal--rich populations.  The colors and
symbols are the same as those in Figure \ref{f12}.  \emph{Bottom:} these panels
plot the [$\alpha$/Fe] ratios as a function of [Na/Fe] (left) and [Al/Fe]
(right).  Note that all of the ``low--$\alpha$" stars have low [Na/Fe] and
[Al/Fe] abundances.}
\label{f13}
\end{figure}

In Figure \ref{f13} and Table 8, we compare the [Na/Fe], [Mg/Fe], [Al/Fe], and 
[$\alpha$/Fe] abundances of the three different metallicity groups in NGC 6273.
Similar to the results of Johnson et al. (2015a), we find that both [Na/Fe] and
[Al/Fe] vary by about factors of 5 and 10, respectively, and that clear Na--Al 
correlations are independently present in the metal--poor, metal--intermediate,
and metal--rich populations.  Therefore, NGC 6273 shares a common feature 
observed in other iron--complex clusters: each population with a unique 
metallicity was able to generate its own independent spread of light element 
abundances that closely resembles the patterns exhibited by monometallic 
clusters.

The Na--Al correlation in Figure \ref{f13} shows a paucity of stars near 
[Al/Fe] $\sim$ 0.6 dex.  If we adopt this cut--off as the discriminator between 
first and second generation stars, then we find that approximately two--thirds 
of the cluster stars can be classified as second generation stars.  The 
metal--poor and metal--intermediate populations each favor second generation 
stars with first:second generation ratios of 35$\%$:65$\%$ and 28$\%$:72$\%$, 
respectively.  On the other hand, the metal--rich population has a ratio of 
75$\%$:25$\%$, but this measurement is based on only 4 stars.  Therefore, the 
numerical dominance of second generation stars in NGC 6273 fits a common trend 
observed in many Galactic globular clusters (e.g., Carretta et al. 2009a, see 
their Figure 10).  Similarly, Figure \ref{f9} shows that the second generation 
stars in NGC 6273 are more centrally concentrated than the first 
generation stars, which again matches a pattern observed in many iron--complex 
and monometallic clusters (e.g., Lardo et al. 2011).  We note also that an 
additional paucity of stars may be present near [Na/Fe] $\sim$ 0.35 dex, which 
may further distinguish the most Na/Al--rich stars.  These stars, which 
constitute $\sim$ 33$\%$ of our sample, are likely equivalent to the ``extreme" 
population found by Carretta et al. (2009a) in several clusters, the ``E" 
population of NGC 2808 (Milone et al. 2015b), and the faint sub--giant branch 
stars of 47 Tuc (Marino et al. 2016).  We note that a similarly high fraction 
of very Na/Al--rich stars is also found in $\omega$ Cen (Johnson \& Pilachowski
2010; Marino et al. 2011) and M 54 (Carretta et al. 2010a).

For [Mg/Fe] and [Al/Fe], Figure \ref{f13} shows that the behavior of the 
element pair may change for stars of different metallicity in NGC 6273.  The 
metal--poor component shows no correlation between [Mg/Fe] and [Al/Fe], but 
the metal--intermediate population shows evidence of a Mg--Al anti--correlation
for stars with [Al/Fe] $<$ 1.0 dex.  A similar Mg--Al anti--correlation may 
also be present for the metal--rich stars, but the sample size (4 stars) is too
small to draw any clear conclusions.  Since \iso{24}{Mg} is only significantly 
depleted at temperatures $\ga$ 65 MK (e.g., see Prantzos et al. 2007; their 
Figure 2), the different Mg--Al relations for the metal--poor and 
metal--intermediate stars suggest that the gas from which each population's 
second generation stars formed was processed at different temperatures.  
However, we did not find any residual correlations between Mg or Al and the 
heavier elements like Si, which indicates that the pollution source(s) 
responsible for the Mg--Al anti--correlation in NGC 6273 likely did not reach 
temperatures high enough to significantly activate the 
\iso{27}{Al}(p,$\gamma$)\iso{28}{Si} reaction related to the Mg--Al chain.

An examination of Mg--Al trends in the iron--complex clusters $\omega$ Cen, 
M 54, M 2, M 22, NGC 1851, and NGC 5286 revealed that only $\omega$ Cen (Norris
\& Da Costa 1995; Smith et al. 2000; Da Costa et al. 2013), M 54 (Carretta 
et al. 2010a), and NGC 1851 (Carretta et al. 2011, 2012) exhibit evidence of 
Mg--Al anti--correlations.  Unlike NGC 6273, none of these clusters show 
evidence that the presence of a Mg--Al anti--correlation depends on a 
population's metallicity.  However, we note that in $\omega$ Cen the 
metal--intermediate and metal--rich stars exhibit clear changes in their light
element patterns (e.g., Norris \& Da Costa 1995; Johnson \& Pilachowski 2010; 
Marino et al. 2011).  For example, the very O--poor/Na--rich stars that 
dominate by number at higher metallicity are not found at [Fe/H] $\la$ --1.8, 
and O and Na are actually correlated in the most metal--rich stars.  
Additionally, for M 54 Carretta et al. (2010a) found that the light element 
variations are more extended for the metal--rich stars than the metal--poor
population.  Therefore, NGC 6273, $\omega$ Cen, and M 54 provide evidence that 
a cluster's enrichment signature can change with time, and that multiple 
pollution sources may be able to produce chemical patterns that are similar 
for some element pairs (e.g., Na--Al) but not others (e.g., Mg--Al).

Interestingly, Figure \ref{f13} shows that the metal--intermediate stars with 
[Al/Fe] $>$ 1.0 dex have [Mg/Fe] $\sim$ 0.4 dex, rather than the [Mg/Fe] $\sim$
0.0 dex abundances that might be expected.  The reason for this discrepancy is 
not immediately clear, but we note that many similar metallicity clusters have 
stars with [Mg/Fe] $\sim$ 0.4 dex and [Al/Fe] $\sim$ 1.0 dex (e.g., Carretta et
al. 2009b; see their Figure 6).  Additionally, we note that Norris \& Da Costa 
(1995) found that intermediate metallicity stars in $\omega$ Cen could have 
[Al/Fe] $>$ 1.0 dex but [Mg/Fe] could range from about 0.6 dex (no Mg--Al
anti--correlation) to 0.0 dex (clear Mg--Al anti--correlation; see also
Carretta et al. 2010a, their Figure 18).  In this context, an examination of 
the \iso{24}{Mg}, \iso{25}{Mg}, and \iso{26}{Mg} abundances in NGC 6273, 
similar to the analysis of Da Costa et al. (2013) in $\omega$ Cen, could 
be particularly illuminating.  However, we also caution that the 6319 \AA\ 
\ion{Mg}{1} lines used here are relatively weak, especially in stars with 
intrinsically low [Mg/Fe], so the exact shape of the Mg--Al anti--correlation 
should be confirmed with additional analyses.

Finally, we note that all of the low--$\alpha$ stars have [Na/Fe] and [Al/Fe] 
compositions that are consistent with those of first generation stars.  
Although the sample size of low--$\alpha$ stars is small, a simulation of 
10$^{\rm 5}$ random draws from our $\alpha$--enhanced population indicated that
there is only about a 0.05$\%$ chance that we would randomly draw 5 stars
that have [Na/Fe] $<$ 0.25 dex and [Al/Fe] $<$ 0.50 dex.  Therefore, we 
speculate that the low--$\alpha$ population may have been unable to form
second generation stars.  The pattern of low [Na/Fe] and [Al/Fe] abundances 
in the low--$\alpha$ stars of NGC 6273 mirrors the composition differences 
found by Carretta et al. (2010a) when comparing the M 54 cluster and 
Sagittarius galaxy field stars.  The similar composition patterns of the
low--$\alpha$ stars in NGC 6273 and the Sagittarius field stars strengthens the 
idea that NGC 6273 may have accreted its low--$\alpha$ stars from a surrounding
field population that was once part of a now dispersed dwarf galaxy.

\subsubsection{RGB versus AGB Abundance Patterns}

As noted by Gratton et al. (2010) and many previous authors (e.g., Mallia 
1978; Norris et al. 1981; Suntzeff 1981; Smith \& Norris 1993; Pilachowski et 
al. 1996a; Ivans et al. 1999; Sneden et al. 2000), some globular clusters may
contain RGB and AGB populations with different light element abundances.
Specifically, RGB stars that evolve onto the HB with masses $\la$ 0.55 
M$_{\rm \odot}$, presumably those with the highest He, N, Na, and Al abundances
and lowest C, O, and Mg abundances, may not ascend the AGB and instead end 
their lives as AGB--manqu{\'e} stars (e.g., Greggio \& Renzini 1990).  As a 
result, we expect to find that the light element abundance distributions of 
AGB stars should exhibit a paucity of second generation stars when compared
to the RGB ratios.

Renewed interest in this field has produced somewhat conflicting results with
the missing second generation fraction ranging from 100$\%$ (Campbell et al.
2013; MacLean et al. 2016) to only a few percent (Johnson \& Pilachowski 2012;
Garc{\'{\i}}a--Hern{\'a}ndez et al. 2015; Johnson et al. 2015b; Lapenna et al.
2016; Wang et al. 2016).  However, the growing consensus is that only the 
most extreme second generation stars probably fail to ascend the AGB.  

\begin{figure}
\epsscale{1.00}
\plotone{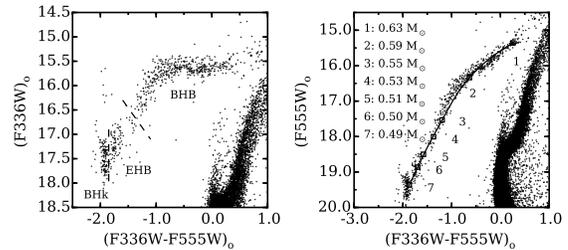}
\caption{\emph{left:} an (F336W)$_{\rm o}$ versus (F336W--F555W)$_{\rm o}$
color--magnitude diagram is shown for the HB region.  The black dashed lines
approximately separate the blue HB (BHB), extreme HB (EHB), and blue hook
(BHk) stellar populations.  Note that the EHB may be composed of at least
three subgroups that each span $\sim$0.5 magnitudes in (F336W)$_{\rm o}$
but $\la$ 0.05--0.10 magnitudes in (F336W--F555W)$_{\rm o}$ color.  A similarly
small color spread is exhibited by the BHk stars as well.  The small color
ranges suggest mass ranges of $\la$ 0.01 M$_{\rm \odot}$ (e.g., see also
Sosin et al. 1997; Momany et al. 2004).  \emph{right:} a similar
(F555W)$_{\rm o}$ versus (F336W--F555W)$_{\rm o}$ color--magnitude diagram is
shown that includes a 12 Gyr, $\alpha$--enhanced BASTI (Pietrinferni et al.
2006) isochrone of [Fe/H] = --1.62 (solid black line).  The isochrone assumes a
cluster distance of 9 kpc and has been shifted by --0.12 magnitudes in color in
order to fit the red end of the blue HB.  The open black boxes labeled 1--7
correspond to temperatures of 8,000 K, 11,500 K, 16,675 K, 20,000 K, 22,500 K,
25,800 K, and 32,000 K, respectively.}
\label{f14}
\end{figure}

\begin{figure}
\epsscale{1.00}
\plotone{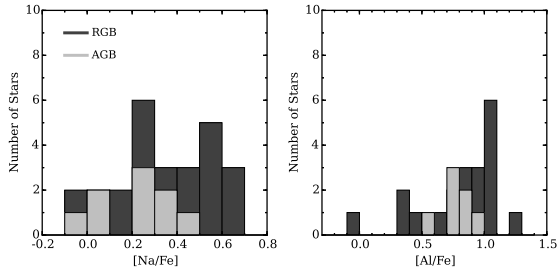}
\caption{The left and right panels compare the [Na/Fe] and [Al/Fe]
distributions of the RGB (dark grey) and AGB (light grey) populations seen
in Figure \ref{f11}.  Note that the AGB stars span a smaller range in both
[Na/Fe] and [Al/Fe], and that no AGB stars were observed to have
[Na/Fe] $>$ 0.50 dex and [Al/Fe] $>$ 1.0 dex.}
\label{f15}
\end{figure}

Since Figure \ref{f14} shows that NGC 6273 contains a very extended blue HB, 
and that $\sim$30$\%$ of the cluster's HB stars have masses $\la$ 0.55 
M$_{\rm \odot}$, we investigate here whether any second generation stars may 
have failed to ascend the AGB.  We restrict the comparison to only the targets 
shown in Figure \ref{f11} since these are the only stars in our sample that 
can be reliably assigned to either the RGB or AGB sequences.  Although the 
sample sizes are small (9 AGB; 28 RGB), we find similar first:second generation
ratios of 24$\%$:76$\%$ and 14$\%$:86$\%$ for the RGB and AGB samples, 
respectively.  However, further inspection of the [Na/Fe] and [Al/Fe] 
distributions in Figure \ref{f15} reveals that the AGB sample does not contain 
stars with [Na/Fe] $>$ 0.5 dex nor [Al/Fe] $>$ 1.0 dex.  In other words, only 
the most Na/Al--rich, and presumably He--enhanced, stars may have failed to
ascend the AGB.  

It is possible that the paucity of extreme Na/Al--rich AGB stars is a product
of our small sample size.  To investigate this, we performed 10$^{\rm 5}$ 
random draws of an equivalent AGB sample from the RGB distribution, and we 
found about a 7$\%$ chance that the missing Na/Al--rich AGB stars could be due
to the small sample size.  Interestingly, the missing Na/Al--rich AGB stars 
account for $\sim$30$\%$ of the RGB sample, which is comparable to the fraction
of extreme HB and blue hook stars found on the HB (see Figure \ref{f14} and 
Section 5.4).  Therefore, we conclude that the NGC 6273 RGB stars with [Na/Fe] 
$>$ 0.5 dex and [Al/Fe] $>$ 1.0 dex likely evolve to become extreme HB or blue 
hook stars and fail to ascend the AGB.

\subsubsection{Fe--Peak Element Abundances}

The Fe--peak elements Cr and Ni are largely produced in the late burning 
stages of massive stars, but include some production by Type Ia SNe as well
(e.g., Timmes et al. 1995).  Within a single globular cluster, the 
star--to--star scatter in [Ni/Fe] and [Cr/Fe] is typically $\la$ 0.1 dex
(e.g., Gratton et al. 2004; see their Figure 2).  Similarly, the average 
[Cr/Fe] and [Ni/Fe] ratios are about solar across the entire metallicity range
spanned by clusters in the Galaxy.

\begin{figure}
\epsscale{1.00}
\plotone{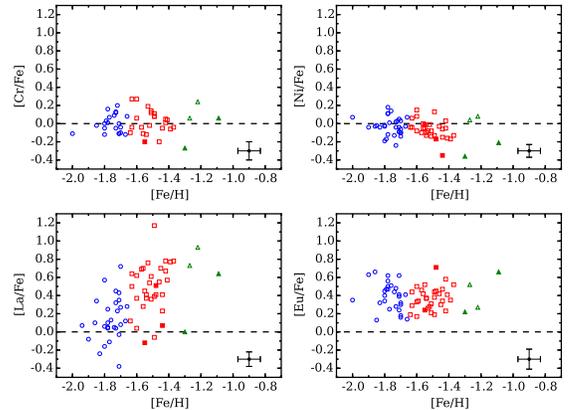}
\caption{The [Cr/Fe], [Ni/Fe], [La/Fe], and [Eu/Fe] abundances are plotted as
a function of [Fe/H] for all three major populations in NGC 6273.  The colors
and symbols are the same as those in Figure \ref{f12}.}
\label{f16}
\end{figure}

In Figure \ref{f16} and Table 8, we show the abundance patterns of [Cr/Fe] and
[Ni/Fe] for NGC 6273.  Overall, we find $\langle$[Cr/Fe]$\rangle$ = 0.01 dex 
($\sigma$ = 0.12 dex) and $\langle$[Ni/Fe]$\rangle$ = --0.05 dex ($\sigma$ = 
0.11 dex), which is in agreement with Johnson et al. (2015a).  An examination
of Figure \ref{f16} shows that the metal--poor, metal--intermediate, and 
metal--rich stars all exhibit nearly identical [Cr/Fe] and [Ni/Fe] abundances
and dispersions.  However, we note that several (but not all) of the 
low--$\alpha$ stars have [Cr,Ni/Fe] $\la$ --0.2 dex, similar to what is found
in some clusters associated with the Sagittarius dwarf galaxy.  A detailed 
examination of key Fe--peak elements that are sensitive to nucleosynthesis 
processes operating in different environments, such as Mn, Co, Zn, and Cu 
(e.g., Nomoto et al. 2006), may provide additional insight into whether the 
stars with low [$\alpha$/Fe], [Cr/Fe], and [Ni/Fe] have similar origins.

\subsubsection{Neutron--Capture Element Abundances}

Most of the stable isotopes heavier than the Fe--peak are produced either 
by the r--process over short time scales or by the s--process over much longer
time scales (e.g., see review by Sneden et al. 2008).  As a result, old
globular clusters tend to have heavy element compositions that are dominated by
r--process nucleosynthesis, which is evidenced by their characteristically low 
[La/Eu] ratios (e.g., see Gratton et al. 2004; their Figure 6).  Although the
Galactic globular cluster system exhibits a trend of increasing s--process
contributions at higher [Fe/H] (e.g., James et al. 2004), a small number of
clusters, such as M4 (e.g., Ivans et al. 1999), deviate from this trend and
exhibit significantly higher [La/Eu] ratios.  In these clusters, the gas from
which their stars formed likely experienced additional, but uniform, pollution 
from a previous generation of $\sim$1.5--4 M$_{\rm \odot}$ AGB stars (e.g., 
Busso et al. 1999).

As mentioned in Section 1, one of the ``chemical tags" of iron--complex 
clusters is that they exhibit clear correlations between [Fe/H] and the 
products of s--process enrichment.  All iron--complex clusters for which the 
heavy elements have been measured contain populations of Fe/s--poor and 
Fe/s--rich stars with similar Ba and La enhancements (Marino et al. 2015; 
Johnson et al. 2015a).  As a result, merger scenarios seem unlikely for every 
case because each cluster would have had to form from the coalescence of 
populations with nearly identical Fe/s--poor and Fe/s--rich compositions (but
see also Gavagnin et al. 2016).  Instead, we regard the combination of [Fe/H] 
and s--process enhancements as a sign that iron--complex clusters were able to 
sustain extended star formation and self--enrichment, and that the time frame 
was long enough for low and intermediate mass AGB stars to contribute to the 
composition of the more metal--rich stars.

\begin{figure}
\epsscale{1.00}
\plotone{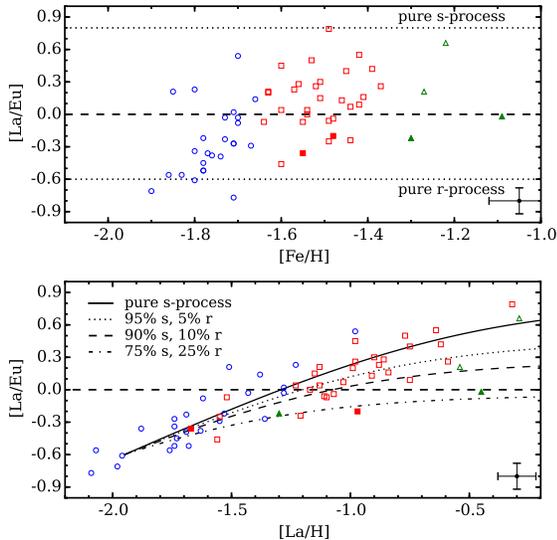}
\caption{\emph{Top:} this panel shows the correlation between [La/Eu] and
[Fe/H] for all NGC 6273 stars observed in this work and Johnson et al. (2015a).
The colors and symbols are the same as in Figure \ref{f12}.  The dotted lines
indicate the pure r--process and pure s--process [La/Eu] abundances from
Kappeler et al. (1989) and Bisterzo et al. (2010), respectively.
\emph{Bottom:} similar to Figure 10 in Johnson et al. (2015a) and following
McWilliam et al. (2013), this panel plots [La/Eu] as a function of [La/H].  The
solid black line indicates the expected change in [La/Eu] as a function of
[La/H] when pure s--process material is added to an initial composition of pure
r--process material.  The dotted, dashed, and dot--dashed dilution curves
represent constant mixtures of 95$\%$(s)/5$\%$(r), 90$\%$(s)/10$\%$(r), and
75$\%$(s)/25$\%$(r) material added to an initial r--process composition.  Note
that the ``low--$\alpha$" stars tend to have low [La/Eu] ratios compared to
stars with similar [La/H] or [Fe/H].}
\label{f17}
\end{figure}

Figure \ref{f16} and Table 8 show a clear increase in [La/Fe] with [Fe/H] for
NGC 6273, in agreement with the results of Johnson et al. (2015a).  Therefore,
we confirm that NGC 6273 possesses the same s--process enrichment profiles as
other iron--complex clusters.  We also find for Eu that the cluster average 
is about [Eu/Fe] = 0.4 dex, regardless of a star's metallicity.  This suggests 
that massive stars were largely responsible for the increase in [Fe/H] within 
the cluster, and that the production rate of Fe and Eu was approximately 
constant.  In Figure \ref{f17}, we update the analysis of Johnson et al. 
(2015a) with a sample size that is $\sim$3$\times$ larger and confirm that the
rise in [La/Fe], and thus the [La/Eu] ratio, with metallicity is due to almost
pure s--process enrichment.  In fact, if we assume that the most La--poor stars 
represent the initial pure r--process composition of the cluster, a simple 
dilution model shows that nearly all of the stars can be accounted for by 
adding $\sim$90$\%$ s--process material and $\sim$10$\%$ r--process material to
the initial r--process composition.  The constant r--process contribution is 
qualitatively in agreement with the [Eu/Fe] observations of Figure \ref{f16} 
because some level of r--process enrichment is required to maintain the 
cluster's overall Eu enhancement at higher [Fe/H].

Interestingly, the low--$\alpha$ stars in Figure \ref{f16} either have [La/Fe] 
$\sim$ 0.6 dex and [Eu/Fe] $\sim$ 0.7 dex or [La/Fe] $\sim$ 0.0 dex and [Eu/Fe]
$\sim$ 0.2 dex.  Although the origin of these stars is not clear, it is 
tempting to speculate that two different formation channels may exist (e.g.,
\emph{in situ} versus accretion).  We note in particular that low--$\alpha$ 
stars with high [La/Fe] and [Eu/Fe] are found in the Sagittarius field (e.g., 
McWilliam et al. 2013), albeit at higher [Fe/H].  The existence of these stars 
further strengthens the idea that at least some of the low--$\alpha$ stars in 
NGC 6273 could have been accreted from a surrounding field population.  The 
low--$\alpha$ stars with lower [La/Fe] and [Eu/Fe] are perhaps a bigger puzzle,
but they could have been formed \emph{in situ} and preferentially enriched by 
Type Ia SNe or massive stars with peculiar enrichment signatures.  However, 
Figure \ref{f17} shows that all of the low--$\alpha$ stars have about the same 
[La/Eu] ratios, and may even fall on a separate enrichment sequence.  In any 
case, the simple dilution model shown in Figure \ref{f17} suggests that the 
low--$\alpha$ stars experienced significant r--process enrichment compared 
to a majority of the $\alpha$--enhanced metal--intermediate and metal--rich 
cluster stars.

\subsection{A Connection Between Blue Hook Stars and Cluster Formation?}

NGC 6273 has long been known to exhibit a peculiar HB morphology that includes 
a very extended blue HB, a clear gap near temperatures of $\sim$20,000 K, and
a large population of blue hook stars (Piotto et al. 1999; Brown et al. 2001, 
2010; Momany et al. 2004).  We confirm these features with new \emph{HST} 
color--magnitude diagrams in Figure \ref{f14}, and find that NGC 6273's HB 
includes several distinct groups\footnote{We adopt the common notation that 
blue HB stars have T$_{\rm eff}$ $\ga$ 8,000 K, extreme HB stars have 20,000 K 
$\la$ T$_{\rm eff}$ $\la$ 32,000 K, and blue hook stars have T$_{\rm eff}$ 
$\ga$ 32,000 K.  In Figure \ref{f14}, the Grundahl jump (Grundahl et al. 1998, 
1999) and Momany jump (Momany et al. 2002, 2004) are found near 
(F336W--F555W)$_{\rm o}$ $\sim$ --0.5 and --1.25 magnitudes, respectively.}.
Although a detailed examination of each HB group is beyond the scope of this 
paper, we draw attention to NGC 6273's large blue hook population in the 
context of its complex formation history.

\begin{figure}
\epsscale{1.00}
\plotone{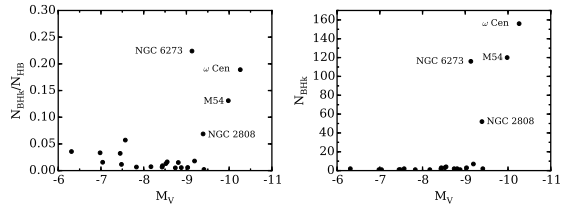}
\caption{\emph{Left:} the ratio of blue hook (BHk) to total HB stars for
several Galactic globular clusters is plotted as a function of absolute
magnitude (M$_{\rm V}$).  Except for NGC 6273, the data are taken from Table 1
of Brown et al. (2016).  Note that NGC 2419 is not shown, but is known to
host a large number of BHk stars as well (e.g., Dieball et al. 2009).
\emph{Right:} the raw number of detected BHk stars is plotted as a function of
absolute magnitude for the same cluster sample.  Both panels indicate that
high masses and large populations of BHk stars distinguish NGC 6273, $\omega$
Cen, M 54, and NGC 2808 from most clusters in the Galaxy.}
\label{f18}
\end{figure}

Blue hook stars are among the hottest core He burning stars in old globular 
clusters, and are thought to form when stars reach the RGB--tip with masses low
enough to delay the core He flash until after a star reaches the white dwarf 
cooling sequence (e.g., D'Cruz et al. 1996; Moehler et al. 2000; Brown et al. 
2010).  The presence of blue hook stars is known to correlate with cluster mass
(Rosenberg et al. 2004; Dieball et al. 2009; Brown et al. 2010, 2016), which 
we illustrate in Figure \ref{f18} by showing that both the ratio of blue hook 
to canonical HB stars ($\frac{N_{BHk}}{N_{HB}}$) and the raw number of blue 
hook stars (N$_{\rm BHk}$) is higher in the more massive clusters.  However, He
enhancement is also likely tied to blue hook formation (e.g., D'Antona et al. 
2002; Tailo et al. 2015).

Although present day cluster mass and the level of He--enrichment strongly 
correlate with the presence of blue hook stars, neither parameter nor a 
combination of the two parameters seems adequate to completely predict blue hook
formation.  For example, He enrichment scenarios (e.g., D'Antona et al. 2010) 
are presently unable to explain the significant carbon enhancements that are 
found in He--enhanced blue hook stars (Moehler et al. 2007, 2011; Latour et al.
2014), and Figure \ref{f18} shows that clusters with similar absolute 
magnitudes (proxies for masses) can have vastly different blue hook 
populations.  To illustrate this point, we note that the iron--complex clusters
NGC 6273 and M 2 differ by only 0.1 magnitudes in M$_{\rm V}$, have similarly 
extended blue HB morphologies, exhibit comparable light and heavy element 
abundance variations, have similar average metallicities and ages, and have 
total HB counts that agree to within 0.5$\%$, but M 2 has a 
$\frac{N_{BHk}}{N_{HB}}$ ratio of 0.006 (3 blue hook stars) whereas NGC 6273 
has $\frac{N_{BHk}}{N_{HB}}$ = 0.224 ($\sim$ 120 blue hook stars)\footnote{The 
HB and blue hook data for M 2 are from Brown et al. (2016).}.  Furthermore, 
dynamical and binary star evolutionary processes may be ruled out as 
explanations because the blue hook stars in clusters with large 
$\frac{N_{BHk}}{N_{HB}}$ ratios, including NGC 6273, do not exhibit radial 
gradients (e.g., Bedin et al. 2000; Brown et al. 2010).  Therefore, additional 
parameters must play a role in producing blue hook stars.

Interestingly, the three objects in Figure \ref{f18} that contain $>$ 100 blue
hook stars and have $\frac{N_{BHk}}{N_{HB}}$ $>$ 0.10 are the iron--complex 
clusters NGC 6273, M 54, and $\omega$ Cen.  All three clusters have about the
same average metallicity, have large metallicity spreads, and exhibit 
extreme variations in light element, heavy element, and (most likely) He
abundances.  However, at least $\omega$ Cen and M 54 are particularly noteworthy
because these clusters are strongly suspected to have extragalactic origins
(e.g., Bekki \& Freeman 2003; Mackey \& van den Bergh 2005).  The similar
chemical pattern and HB morphology that NGC 6273 shares with $\omega$ Cen and
M 54 suggest that NGC 6273 may have also been accreted by the Milky Way.  If
these clusters are all remnants of dwarf galaxy systems, then it is 
reasonable to assume that each cluster has experienced significant mass loss.
Therefore, a cluster's formation environment and initial mass may play
critical roles in forming large populations of blue hook stars, and the 
different blue hook populations of NGC 6273 and M 2 could be explained if NGC 
6273 was initially much more massive than M 2 and/or formed in a different 
environment.  In this context, we note that NGC 2419\footnote{NGC 2419 is 
omitted from Figure \ref{f18} because it was not included in the compilation 
by Brown et al. (2016), but likely also has $>$ 100 blue hook stars (Dieball 
et al. 2009).} and NGC 2808 would also be candidates that may have formed with 
much larger initial masses, and their higher and lower $\frac{N_{BHk}}{N_{HB}}$
ratios compared to NGC 6273 could be driven by their lower and higher 
respective metallicities.  At least for NGC 2419, there are also some 
indications that the cluster may have an extragalactic origin (Mackey \& van 
den Bergh 2005).

\section{SUMMARY}

We have measured detailed abundances, CaT metallicities, and/or radial 
velocities for $>$ 800 RGB stars ($>$ 300 members) near the massive bulge 
globular cluster NGC 6273.  The abundances and velocities are based on an 
analysis of high resolution (R $\approx$ 27,000) spectra obtained with the 
\emph{Magellan}--M2FS multi--fiber instrument, and includes additional 
metallicity and velocity measurements of R $\approx$ 18,000 archival 
\emph{VLT}--FLAMES CaT spectra.  The new data extend the spectroscopic work of 
Johnson et al. (2015a) and Yong et al. (2016) and span a broad range in 
luminosity and color.  These data are complemented by photometric measurements 
of new \emph{HST}--WFC3/UVIS data in the F336W, F438W, F555W, and F814W bands 
that extend from the RGB--tip down to at least 2 magnitudes below the 
main--sequence turn--off.  

A simple kinematic analysis indicates that $\sim$40$\%$ of our spectroscopic
targets are cluster members and have heliocentric radial velocities between 
$+$120 and $+$170 km s$^{\rm -1}$.  We find a cluster average velocity of 
$+$144.71 km s$^{\rm -1}$ and a dispersion of 8.57 km s$^{\rm -1}$.  The 
cluster exhibits net rotation with a mean projected amplitude of 
3.83 km s$^{\rm -1}$.  A Plummer model fit to the projected radial velocity 
dispersion profile suggests that NGC 6273 has a central velocity dispersion of 
at least 10--12 km s$^{\rm -1}$ and an A$_{\rm rot.}$/$\sigma$$_{\rm o}$ ratio 
of $\sim$0.30--0.35.

The [Fe/H] abundances presented here follow the results of Johnson et al.
(2015a), Han et al. (2015), and Yong et al. (2016) that suggest an intrinsic
metallicity spread exists in NGC 6273.  Using EW measurements of individual 
\ion{Fe}{1} and \ion{Fe}{2} lines, we find evidence that at least three 
stellar populations with different [Fe/H] may exist: (1) a metal--poor group 
with [Fe/H] $\leq$ --1.65; (2) a metal--intermediate group with --1.65 $<$ 
[Fe/H] $\leq$ --1.35; and (3) a metal--rich group with [Fe/H] $>$ --1.35.  The 
metal--poor and metal--intermediate populations may be associated with 
different giant branches, and both populations may contain roughly equivalent 
numbers of stars.  In contrast, the metal--rich population only constitutes 
6$\%$ of our sample.  The metal--intermediate stars may also be more centrally 
concentrated than the metal--poor stars, but the radial distribution 
differences are only observed at projected distances $\ga $1.5$\arcmin$ from 
the cluster center.  Similar to Yong et al. (2016), our CaT measurements extend
the metal--rich tail to at least [Fe/H] = --1.0 to --0.5 dex, but it is 
possible that some (or all) of these comparatively very metal--rich stars could
be bulge field stars with velocities in the membership range.

The cluster's chemical abundance patterns indicate that all three major 
populations contain distinct sets of first (Na/Al--poor) and second 
(Na/Al--rich) generation stars.  All three populations exhibit similar 
Na--Al correlations, but the [Mg/Fe] and [Al/Fe] distributions suggest a 
complex enrichment scenario.  For example, [Al/Fe] spans about a factor of 10
in abundance for the metal--poor and metal--intermediate populations, but only 
the metal--intermediate stars show evidence of a Mg--Al anti--correlation.
The metal--rich stars may also exhibit a Mg--Al anti--correlation, but the 
sample size is too small to draw any strong conclusions.  In confirmed, a 
change in the Mg--Al distribution as a function of metallicity may suggest
that the gas from which the metal--intermediate and metal--rich second
generation stars formed was processed at higher temperatures than the gas from
which the metal--poor second generation stars formed.  Notably, we did not 
observe any significant correlations between Mg/Al and Si that would have 
indicated burning temperatures significantly higher than $\sim$65--70 MK, as 
is the case in several other massive clusters.  Interestingly, the 
metal--intermediate and metal--rich stars with [Al/Fe] $>$ 1.0 dex have higher 
than expected [Mg/Fe] abundances, which could indicate that the gas from which 
these stars formed was polluted by a different class or mass range of objects.

Further examination of the light element abundances indicates that the RGB and 
AGB stars may not have identical [Na/Fe] and [Al/Fe] distributions.  In 
particular, we did not find any AGB stars with [Na/Fe] $>$ 0.5 dex or 
[Al/Fe] $>$ 1.0 dex.  The ``missing" AGB stars account for $\sim$30$\%$ of the
RGB sample, which is close to the fraction of extreme HB and blue hook
stars relative to the total HB population.  We speculate that the RGB stars 
with the highest [Na/Fe] and [Al/Fe] abundances likely evolve to become extreme
HB or blue hook stars and do not ascend the AGB.

The overall [$\alpha$/Fe] ratios may slowly decline with increasing 
metallicity, but most stars have [$\alpha$/Fe] $\sim$ 0.3 dex.  Additionally, 
the Fe--peak elements exhibit solar [X/Fe] ratios, regardless of 
metallicity, and the star--to--star dispersion is $\sim$0.1 dex in all three 
populations.  In contrast, the heavy s--process element La exhibits a 
correlated increase with metallicity that ranges from [La/Fe] $\sim$ --0.2 dex 
at the lowest metallicities to [La/Fe] $\sim$ 0.8 dex at the highest 
metallicities.  However, the r--process element Eu maintains a constant 
abundance of [Eu/Fe] $\sim$ 0.4 dex across the full [Fe/H] range.  In agreement
with Johnson et al. (2015a), we find that the correlated increase in [La/Eu] 
with metallicity is consistent with a nearly pure s--process enrichment 
pattern.  Constant r--process production is required to maintain the flat 
[Eu/Fe] abundance distribution, but the r--process contribution likely does 
not significantly exceed $\sim$10$\%$.  Therefore, we confirm that NGC 6273 
shares an almost identical s--process enrichment pattern with other 
iron--complex clusters such as $\omega$ Cen, M 2, M 22, and NGC 5286.

We have also identified a population of at least 5 peculiar ``low--$\alpha$" 
stars that have [$\alpha$/Fe] $\sim$ 0.0 dex, low [Na/Fe] and [Al/Fe] abundances
(all are first generation stars), and low [La/Eu] ratios.  Many, but not all, 
of the low--$\alpha$ stars also exhibit low [Cr/Fe] and [Ni/Fe] abundances.  
Although the metal--poor population does not contain any low--$\alpha$ stars, 
the specific frequency of low--$\alpha$ stars increases from 9$\%$ in the 
metal--intermediate population to 50$\%$ in the metal--rich population.  
However, the ratios of low--$\alpha$ stars in each population should be 
confirmed with future large sample observations.  Interestingly, the 
combination of $\alpha$--enhanced and $\alpha$--poor stars in NGC 6273 closely 
resembles the M 54 and Sagittarius field star system, and we speculate that 
some or all of the low--$\alpha$ stars may have been accreted from a former 
field population that surrounded the NGC 6273 core but had different 
chemistry.  We note that similar populations may also be present in at 
least $\omega$ Cen, M 2, and NGC 5286.

An examination of NGC 6273's HB revealed a particularly complex morphology.
We find that the HB is composed of several distinct groups of stars with 
different masses, and that the mass range within each of the extreme HB and
blue hook populations varies by $\la$ 0.01 M$_{\rm \odot}$.  Interestingly, the 
\emph{HST} data show that NGC 6273 may have one of the largest blue hook 
populations in the Galaxy.  In particular, the ratio of blue hook to canonical 
HB stars is $\sim$0.22, which is a trait shared only by $\omega$ Cen, M 54, 
NGC 2419, and NGC 2808.  Since all of these clusters are very massive, and at 
least $\omega$ Cen, M 54, and NGC 2419 are suspected to have extragalactic 
origins, we speculate that a cluster's initial mass and formation environment 
are likely critical factors in the production of blue hook stars, at least in 
large numbers.

\acknowledgements

This research has made use of NASA's Astrophysics Data System Bibliographic
Services.  This publication has made use of data products from the Two Micron
All Sky Survey, which is a joint project of the University of Massachusetts
and the Infrared Processing and Analysis Center/California Institute of
Technology, funded by the National Aeronautics and Space Administration and
the National Science Foundation.  C.I.J. gratefully acknowledges support from
the Clay Fellowship, administered by the Smithsonian Astrophysical Observatory.
M.M. is grateful for support from the National Science Foundation to develop 
M2FS (AST--0923160) and carry out the observations reported here 
(AST--1312997), and to the University of Michigan for its direct support of 
M2FS construction and operation.  M.G.W. is supported by National Science 
Foundation grants AST--1313045 and AST--1412999.  R.M.R acknowledges support 
from grant AST--1413755 from the National Science Foundation.  E.W.O. 
acknowledges support from the National Science Foundation under grant 
AST--1313006.  C.I.J. would like to thank David Yong for providing electronic 
data tables of his work, and Antonino Milone for kindly providing differential 
reddening measurements for this paper.  Support for program \#GO--14197 was 
provided by NASA through a grant from the Space Telescope Science Institute, 
which is operated by the Association of Universities for Research in Astronomy,
Inc., under NASA contract NAS 5--26555.

\clearpage
\LongTables
\begin{landscape}

\tablenum{1}
\tablecolumns{5}
\tablewidth{0pt}

\begin{deluxetable}{ccccc}
\tablecaption{Observing Log}
\tablehead{
\colhead{Field\tablenotemark{a}}        &
\colhead{Telescope/Instrument}  &
\colhead{Setup}      &
\colhead{UT Date}      &
\colhead{Exposure}      \\
\colhead{}      &
\colhead{}      &
\colhead{}      &
\colhead{}      &
\colhead{(s)}
}

\startdata
\hline
\multicolumn{5}{c}{Spectroscopy}       \\
\hline
1a	&	VLT$-$FLAMES	&	HR21	&	2014 April 13	&	1 $\times$ 2445	\\
1b	&	VLT$-$FLAMES	&	HR21	&	2014 May 7	&	1 $\times$ 2445	\\
2a	&	VLT$-$FLAMES	&	HR21	&	2014 July 13	&	1 $\times$ 2445	\\
2b	&	VLT$-$FLAMES	&	HR21	&	2014 August 2	&	1 $\times$ 2445	\\
3a	&	VLT$-$FLAMES	&	HR21	&	2014 July 21	&	1 $\times$ 2445	\\
3b	&	VLT$-$FLAMES	&	HR21	&	2014 August 13	&	1 $\times$ 2445	\\
4	&	Magellan$-$M2FS	&	CaT	&	2015 July 17	&	4 $\times$ 1200	\\
\enddata

\tablenotetext{a}{Fields with different designations indicate different 
telescope pointings.  The ``a" and ``b" designations for the VLT--FLAMES
setups correspond to the ``HIERARCH ESO OBS NAME" keyword in the original
image headers.  The ``a" and ``b" fields with the same numbers typically 
observed the same stars, but sometimes with different fibers.}

\end{deluxetable}

\clearpage

\tablenum{2}
\tablecolumns{7}
\tablewidth{0pt}

\begin{deluxetable}{ccccccc}
\tablecaption{Star Identifiers, Coordinates, Photometry, and Radial Velocities
for NGC 6273 Members}
\tablehead{
\colhead{Star Name}	&
\colhead{RA}	&
\colhead{DEC}      &
\colhead{J}      &
\colhead{K$_{\rm S}$}	&
\colhead{RV$_{\rm helio.}$}	&
\colhead{RV Error}	\\
\colhead{(2MASS)}	&
\colhead{(degrees)}	&
\colhead{(degrees)}	&
\colhead{(mag.)}	&
\colhead{(mag.)}        &
\colhead{(km s$^{\rm -1}$)}      &
\colhead{(km s$^{\rm -1}$)}      
}

\startdata
\hline
\multicolumn{7}{c}{Bulge$\_$GC1 Members}       \\
\hline
17022227$-$2613433\tablenotemark{d}	&	255.592801	&	$-$26.228718	&	10.882	&	9.920	&	142.87	&	0.16	\\
17022817$-$2616426	&	255.617398	&	$-$26.278500	&	11.893	&	11.102	&	156.50	&	0.23	\\
17022912$-$2617443\tablenotemark{d}	&	255.621349	&	$-$26.295652	&	11.153	&	10.218	&	143.61	&	0.18	\\
17023087$-$2618515	&	255.628646	&	$-$26.314312	&	11.891	&	11.103	&	149.94	&	0.27	\\
17023192$-$2614177\tablenotemark{d}	&	255.633037	&	$-$26.238272	&	10.451	&	9.434	&	125.82	&	0.21	\\
17023225$-$2614521	&	255.634399	&	$-$26.247812	&	11.764	&	10.896	&	123.72	&	0.19	\\
17023338$-$2617104\tablenotemark{d}	&	255.639093	&	$-$26.286234	&	12.075	&	11.262	&	143.75	&	0.37	\\
\enddata

\tablenotetext{a}{Observed in Johnson et al. (2015), the Bulge$\_$GC1 setup,
and the M2FS Calcium Triplet setup.}
\tablenotetext{b}{Observed in Johnson et al. (2015) and the Bulge$\_$GC1
setup.}
\tablenotetext{c}{Observed in Johnson et al. (2015) and the M2FS Calcium 
Triplet setup.}
\tablenotetext{d}{Observed in the Bulge$\_$GC1 and M2FS Calcium Triplet setups.}
\tablenotetext{e}{Observed in the M2FS Calcium Triplet and FLAMES Calcium 
Triplet setups.}

\end{deluxetable}

\clearpage

\tablenum{3}
\tablecolumns{7}
\tablewidth{0pt}

\begin{deluxetable}{ccccccc}
\tablecaption{Star Identifiers, Coordinates, Photometry, and Radial Velocities
for Non--Members}
\tablehead{
\colhead{Star Name}	&
\colhead{RA}	&
\colhead{DEC}      &
\colhead{J}      &
\colhead{K$_{\rm S}$}	&
\colhead{RV$_{\rm helio.}$}	&
\colhead{RV Error}	\\
\colhead{(2MASS)}	&
\colhead{(degrees)}	&
\colhead{(degrees)}	&
\colhead{(mag.)}	&
\colhead{(mag.)}        &
\colhead{(km s$^{\rm -1}$)}      &
\colhead{(km s$^{\rm -1}$)}      
}

\startdata
\hline
\multicolumn{7}{c}{Bulge$\_$GC1 Non--Members}       \\
\hline
17020064$-$2611478	&	255.502696	&	$-$26.196625	&	11.317	&	10.483	&	87.50	&	0.15	\\
17020290$-$2612561	&	255.512110	&	$-$26.215591	&	11.410	&	10.543	&	$-$69.71	&	0.17	\\
17020445$-$2612074	&	255.518548	&	$-$26.202074	&	10.501	&	9.434	&	5.64	&	0.32	\\
17020743$-$2611048	&	255.530968	&	$-$26.184685	&	10.423	&	9.362	&	94.73	&	0.32	\\
17021345$-$2620018	&	255.556074	&	$-$26.333847	&	10.539	&	9.579	&	$-$23.62	&	0.27	\\
17021419$-$2615558\tablenotemark{b}	&	255.559142	&	$-$26.265518	&	11.537	&	10.704	&	$-$14.01	&	0.17	\\
17021609$-$2622447	&	255.567065	&	$-$26.379099	&	11.789	&	10.937	&	65.75	&	0.34	\\
\enddata

\tablenotetext{a}{Observed in Johnson et al. (2015) and the M2FS Calcium 
Triplet setup.}
\tablenotetext{b}{Observed in the Bulge$\_$GC1 and M2FS Calcium Triplet setups.}

\end{deluxetable}

\clearpage

\tablenum{4}
\tablecolumns{5}
\tablewidth{0pt}

\begin{deluxetable}{ccccc}
\tablecaption{Model Atmosphere Parameters for NGC 6273 Members}
\tablehead{
\colhead{Star Name}	&
\colhead{T$_{\rm eff}$}	&
\colhead{log(g)}      &
\colhead{[Fe/H]}      &
\colhead{$\xi$$_{\rm mic.}$}      \\
\colhead{(2MASS)}	&
\colhead{(K)}      &
\colhead{(cgs)}      &
\colhead{(dex)}      &
\colhead{(km s$^{\rm -1}$)} 
}

\startdata
17022227$-$2613433\tablenotemark{c}	&	4400	&	1.15	&	$-$1.67	&	1.90	\\
17022817$-$2616426	&	4675	&	1.75	&	$-$1.66	&	1.95	\\
17022912$-$2617443\tablenotemark{c}	&	4325	&	0.80	&	$-$1.80	&	1.80	\\
17023087$-$2618515	&	4575	&	1.10	&	$-$1.86	&	1.95	\\
17023192$-$2614177\tablenotemark{c}	&	4325	&	1.35	&	$-$1.49	&	2.10	\\
17023225$-$2614521	&	4500	&	1.15	&	$-$1.71	&	1.85	\\
17023338$-$2617104\tablenotemark{c}	&	4675	&	1.50	&	$-$1.63	&	1.85	\\
17023342$-$2616165	&	4575	&	1.65	&	$-$1.51	&	1.75	\\
\enddata

\tablenotetext{a}{Observed in Johnson et al. (2015), the Bulge$\_$GC1 setup,
and the M2FS Calcium Triplet setup.}
\tablenotetext{b}{Observed in Johnson et al. (2015) and the Bulge$\_$GC1
setup.}
\tablenotetext{c}{Observed in the Bulge$\_$GC1 and M2FS Calcium Triplet setups.}

\end{deluxetable}

\clearpage

\tablenum{5}
\tablecolumns{13}
\tablewidth{0pt}

\begin{deluxetable}{ccccccccccccc}
\tabletypesize{\tiny}
\tablecaption{Chemical Abundances for NGC 6273 Members: Na$-$Cr}
\tablehead{
\colhead{Star Name}	&
\colhead{[Na I/Fe]}	&
\colhead{$\Delta$[Na I/Fe]}      &
\colhead{[Mg I/Fe]}       &
\colhead{$\Delta$[Mg I/Fe]}      &
\colhead{[Al I/Fe]}       &
\colhead{$\Delta$[Al I/Fe]}      &
\colhead{[Si I/Fe]}       &
\colhead{$\Delta$[Si I/Fe]}      &
\colhead{[Ca I/Fe]}       &
\colhead{$\Delta$[Ca I/Fe]}      &
\colhead{[Cr I/Fe]}       &
\colhead{$\Delta$[Cr I/Fe]}      \\
\colhead{(2MASS)}	&
\colhead{(dex)}	&
\colhead{(dex)} &
\colhead{(dex)} &
\colhead{(dex)} &
\colhead{(dex)} &
\colhead{(dex)} &
\colhead{(dex)} &
\colhead{(dex)} &
\colhead{(dex)} &
\colhead{(dex)} &
\colhead{(dex)} &
\colhead{(dex)}
}

\startdata
17022227$-$2613433\tablenotemark{c}	&	0.33	&	0.11	&	0.34	&	0.07	&	1.26	&	0.05	&	0.33	&	0.09	&	0.24	&	0.07	&	$-$0.12	&	0.13	\\
17022817$-$2616426	&	0.27	&	0.06	&	0.43	&	0.07	&	0.79	&	0.07	&	0.32	&	0.14	&	0.34	&	0.07	&	0.08	&	0.06	\\
17022912$-$2617443\tablenotemark{c}	&	0.23	&	0.06	&	0.32	&	0.07	&	0.21	&	0.09	&	0.31	&	0.04	&	0.32	&	0.08	&	$-$0.05	&	0.09	\\
17023087$-$2618515	&	0.16	&	0.06	&	0.37	&	0.07	&	0.50	&	0.07	&	\nodata	&	\nodata	&	0.21	&	0.09	&	\nodata	&	\nodata	\\
17023192$-$2614177\tablenotemark{c}	&	0.31	&	0.06	&	0.31	&	0.07	&	0.62	&	0.10	&	0.18	&	0.09	&	0.14	&	0.08	&	0.11	&	0.09	\\
17023225$-$2614521	&	0.10	&	0.05	&	\nodata	&	\nodata	&	0.11	&	0.07	&	0.23	&	0.13	&	0.15	&	0.07	&	$-$0.11	&	0.09	\\
17023338$-$2617104\tablenotemark{c}	&	0.55	&	0.05	&	\nodata	&	\nodata	&	1.01	&	0.07	&	0.27	&	0.10	&	0.29	&	0.08	&	0.27	&	0.09	\\
17023342$-$2616165	&	0.13	&	0.04	&	\nodata	&	\nodata	&	0.57	&	0.07	&	0.26	&	0.10	&	0.19	&	0.09	&	0.14	&	0.05	\\
\enddata

\tablenotetext{a}{Observed in Johnson et al. (2015), the Bulge$\_$GC1 setup,
and the Calcium Triplet setup.}
\tablenotetext{b}{Observed in Johnson et al. (2015) and the Bulge$\_$GC1
setup.}
\tablenotetext{c}{Observed in Johnson et al. (2015) and the Calcium Triplet
setup.}
\end{deluxetable}

\clearpage

\tablenum{6}
\tablecolumns{11}
\tablewidth{0pt}

\begin{deluxetable}{ccccccccccccc}
\tabletypesize{\tiny}
\tablecaption{Chemical Abundances for NGC 6273 Members: Fe$-$Eu}
\tablehead{
\colhead{Star Name}	&
\colhead{[Fe I/H]}	&
\colhead{$\Delta$[Fe I/H]}      &
\colhead{[Fe II/H]}       &
\colhead{$\Delta$[Fe II/H]}      &
\colhead{[Ni I/Fe]}       &
\colhead{$\Delta$[Ni I/Fe]}      &
\colhead{[La II/Fe]}       &
\colhead{$\Delta$[La II/Fe]}      &
\colhead{[Eu II/Fe]}       &
\colhead{$\Delta$[Eu II/Fe]}      \\
\colhead{(2MASS)}	&
\colhead{(dex)}	&
\colhead{(dex)} &
\colhead{(dex)} &
\colhead{(dex)} &
\colhead{(dex)} &
\colhead{(dex)} &
\colhead{(dex)} &
\colhead{(dex)} &
\colhead{(dex)} &
\colhead{(dex)} 
}

\startdata
17022227$-$2613433\tablenotemark{c}	&	$-$1.67	&	0.07	&	$-$1.67	&	0.08	&	$-$0.03	&	0.07	&	0.12	&	0.08	&	0.41	&	0.07	\\
17022817$-$2616426	&	$-$1.66	&	0.07	&	$-$1.66	&	0.08	&	0.07	&	0.07	&	0.28	&	0.08	&	0.14	&	0.10	\\
17022912$-$2617443\tablenotemark{c}	&	$-$1.80	&	0.07	&	$-$1.80	&	0.08	&	$-$0.02	&	0.08	&	0.57	&	0.06	&	0.34	&	0.10	\\
17023087$-$2618515	&	$-$1.85	&	0.07	&	$-$1.86	&	0.06	&	$-$0.03	&	0.07	&	0.10	&	0.08	&	0.66	&	0.10	\\
17023192$-$2614177\tablenotemark{c}	&	$-$1.48	&	0.07	&	$-$1.49	&	0.08	&	$-$0.13	&	0.08	&	1.17	&	0.13	&	0.38	&	0.10	\\
17023225$-$2614521	&	$-$1.71	&	0.07	&	$-$1.71	&	0.07	&	$-$0.09	&	0.07	&	0.35	&	0.08	&	0.62	&	0.10	\\
17023338$-$2617104\tablenotemark{c}	&	$-$1.63	&	0.07	&	$-$1.63	&	0.09	&	0.06	&	0.07	&	0.64	&	0.06	&	0.44	&	0.10	\\
17023342$-$2616165	&	$-$1.51	&	0.07	&	$-$1.50	&	0.08	&	$-$0.135	&	0.06	&	0.36	&	0.08	&	0.21	&	0.10	\\
\enddata

\tablenotetext{a}{Observed in Johnson et al. (2015), the Bulge$\_$GC1 setup,
and the Calcium Triplet setup.}
\tablenotetext{b}{Observed in Johnson et al. (2015) and the Bulge$\_$GC1
setup.}
\tablenotetext{c}{Observed in Johnson et al. (2015) and the Calcium Triplet
setup.}
\end{deluxetable}

\clearpage

\tablenum{7}
\tablecolumns{7}
\tablewidth{0pt}

\begin{deluxetable}{ccccccc}
\tablecaption{Calcium Triplet Metallicity Data}
\tablehead{
\colhead{Star Name}	&
\colhead{EW$_{\rm 8542}$}	&
\colhead{EW$_{\rm 8662}$}	&
\colhead{$\sum$EW}	&
\colhead{W$^{\prime}$}	&
\colhead{[Fe/H]}	&
\colhead{$\Delta$[Fe/H]}	\\
\colhead{(2MASS)}	&
\colhead{(\AA)}      &
\colhead{(\AA)}      &
\colhead{(\AA)}      &
\colhead{(\AA)}      &
\colhead{(dex)}	&
\colhead{(dex)}
}

\startdata
\hline
\multicolumn{7}{c}{M2FS Calcium Triplet Members}	\\
\hline
17015056$-$2616256	&	\nodata	&	\nodata	&	\nodata	&	\nodata	&	\nodata	&	\nodata	\\
17021380$-$2613223	&	1.99	&	1.45	&	3.44	&	3.17	&	$-$2.01	&	0.10	\\
17021778$-$2616058	&	2.11	&	1.47	&	3.58	&	3.21	&	$-$1.99	&	0.12	\\
17022040$-$2616289\tablenotemark{b}	&	2.54	&	1.77	&	4.31	&	3.50	&	$-$1.89	&	0.13	\\
17022227$-$2613433\tablenotemark{c}	&	3.04	&	2.32	&	5.36	&	4.24	&	$-$1.61	&	0.11	\\
17022395$-$2614538\tablenotemark{b}	&	2.63	&	2.02	&	4.65	&	3.90	&	$-$1.74	&	0.11	\\
17022413$-$2619124	&	2.37	&	1.73	&	4.10	&	3.77	&	$-$1.79	&	0.10	\\
\enddata

\tablenotetext{a}{Observed in Johnson et al. (2015), the Bulge$\_$GC1 setup,
and the M2FS Calcium Triplet setup.}
\tablenotetext{b}{Observed in Johnson et al. (2015) and the M2FS Calcium
Triplet setup.}
\tablenotetext{c}{Observed in the Bulge$\_$GC1 and M2FS Calcium Triplet setups.}
\tablenotetext{d}{Observed in the M2FS Calcium Triplet and FLAMES Calcium
Triplet setups.}

\end{deluxetable}

\clearpage

\tablenum{8}
\tablecolumns{11}
\tablewidth{0pt}

\begin{deluxetable}{ccccccccccc}
\tabletypesize{\footnotesize}
\tablecaption{Mean Composition Properties}
\tablehead{
\colhead{Statistic}	&
\colhead{[Fe/H]}	&
\colhead{[Na/Fe]}	&
\colhead{[Mg/Fe]}      &
\colhead{[Al/Fe]}      &
\colhead{[Si/Fe]}      &
\colhead{[Ca/Fe]}      &
\colhead{[Cr/Fe]}      &
\colhead{[Ni/Fe]}      &
\colhead{[La/Fe]}      &
\colhead{[Eu/Fe]}      \\
\colhead{}      &
\colhead{(dex)}      &
\colhead{(dex)}      &
\colhead{(dex)}      &
\colhead{(dex)}      &
\colhead{(dex)}      &
\colhead{(dex)}      &
\colhead{(dex)}      &
\colhead{(dex)}      &
\colhead{(dex)}      &
\colhead{(dex)}      
}

\startdata
\hline
\multicolumn{11}{c}{Metal$-$Poor Population}	\\
\hline
Average	&	$-$1.77	&	0.30	&	0.40	&	0.74	&	0.35	&	0.25	&	0.00	&	$-$0.03	&	0.15	&	0.39	\\
$\sigma$	&	0.08	&	0.19	&	0.08	&	0.34	&	0.10	&	0.09	&	0.10	&	0.10	&	0.24	&	0.15	\\
\hline
\multicolumn{11}{c}{Metal$-$Intermediate Population}    \\
\hline
Average	&	$-$1.51	&	0.26	&	0.34	&	0.75	&	0.23	&	0.22	&	0.02	&	$-$0.06	&	0.47	&	0.36	\\
$\sigma$	&	0.07	&	0.20	&	0.10	&	0.27	&	0.15	&	0.11	&	0.13	&	0.10	&	0.28	&	0.12	\\
\hline
\multicolumn{11}{c}{Metal$-$Rich Population}    \\
\hline
Average	&	$-$1.22	&	0.19	&	0.17	&	0.46	&	0.11	&	0.16	&	0.02	&	$-$0.11	&	0.57	&	0.42	\\
$\sigma$	&	0.09	&	0.34	&	0.16	&	0.41	&	0.21	&	0.21	&	0.21	&	0.21	&	0.40	&	0.21	\\
\enddata

\end{deluxetable}
\end{landscape}


\begin{thebibliography}{}

\bibitem[Alonso-Garc{\'{\i}}a et al.(2012)]{2012AJ....143...70A} Alonso-Garc{\'{\i}}a, J., Mateo, M., Sen, B., et al.\ 2012, \aj, 143, 70

\bibitem[Armandroff \& Da Costa(1991)]{1991AJ....101.1329A} Armandroff, T.~E., \& Da Costa, G.~S.\ 1991, \aj, 101, 1329

\bibitem[Bastian et al.(2015)]{2015MNRAS.449.3333B} Bastian, N., Cabrera-Ziri, I., \& Salaris, M.\ 2015, \mnras, 449, 3333

\bibitem[Bastian \& Lardo(2015)]{2015MNRAS.453..357B} Bastian, N., \& Lardo, C.\ 2015, \mnras, 453, 357

\bibitem[Battaglia et al.(2008)]{2008MNRAS.383..183B} Battaglia, G., Irwin, M., Tolstoy, E., et al.\ 2008, \mnras, 383, 183

\bibitem[Bedin et al.(2000)]{2000A&A...363..159B} Bedin, L.~R., Piotto, G., Zoccali, M., et al.\ 2000, \aap, 363, 159

\bibitem[Bekki \& Freeman(2003)]{2003MNRAS.346L..11B} Bekki, K., \& Freeman, K.~C.\ 2003, \mnras, 346, L11 

\bibitem[Bellazzini et al.(2008)]{2008AJ....136.1147B} Bellazzini, M., Ibata, R.~A., Chapman, S.~C., et al.\ 2008, \aj, 136, 1147

\bibitem[Bellazzini et al.(2012)]{2012A&A...538A..18B} Bellazzini, M., Bragaglia, A., Carretta, E., et al.\ 2012, \aap, 538, A18

\bibitem[Bellini et al.(2009)]{2009A&A...507.1393B} Bellini, A., Piotto, G., Bedin, L.~R., et al.\ 2009, \aap, 507, 1393

\bibitem[Bensby et al.(2013)]{2013A&A...549A.147B} Bensby, T., Yee, J.~C., Feltzing, S., et al.\ 2013, \aap, 549, A147

\bibitem[Bianchini et al.(2013)]{2013ApJ...772...67B} Bianchini, P., Varri, A.~L., Bertin, G., \& Zocchi, A.\ 2013, \apj, 772, 67

\bibitem[Bisterzo et al.(2010)]{2010MNRAS.404.1529B} Bisterzo, S., Gallino,
R., Straniero, O., Cristallo, S., Kappeler, F.\ 2010, \mnras, 404, 1529

\bibitem[Bragaglia et al.(2010a)]{2010ApJ...720L..41B} Bragaglia, A., Carretta, E., Gratton, R.~G., et al.\ 2010a, \apjl, 720, L41

\bibitem[Bragaglia et al.(2010b)]{2010A&A...519A..60B} Bragaglia, A., Carretta, E., Gratton, R., et al.\ 2010b, \aap, 519, A60

\bibitem[Bragaglia et al.(2014)]{2014ApJ...796...68B} Bragaglia, A., Sneden, C., Carretta, E., et al.\ 2014, \apj, 796, 68

\bibitem[Brown et al.(2001)]{2001ApJ...562..368B} Brown, T.~M., Sweigart, A.~V., Lanz, T., Landsman, W.~B., \& Hubeny, I.\ 2001, \apj, 562, 368

\bibitem[Brown et al.(2010)]{2010ApJ...718.1332B} Brown, T.~M., Sweigart, A.~V., Lanz, T., et al.\ 2010, \apj, 718, 1332

\bibitem[Brown et al.(2016)]{2016ApJ...822...44B} Brown, T.~M., Cassisi, S., D'Antona, F., et al.\ 2016, \apj, 822, 44

\bibitem[Busso et al.(1999)]{1999ARA&A..37..239B} Busso, M., Gallino, R., \& Wasserburg, G.~J.\ 1999, \araa, 37, 239

\bibitem[Campbell et al.(2013)]{2013Natur.498..198C} Campbell, S.~W., D'Orazi, V., Yong, D., et al.\ 2013, \nat, 498, 198

\bibitem[Carrera et al.(2007)]{2007AJ....134.1298C} Carrera, R., Gallart, C., Pancino, E., \& Zinn, R.\ 2007, \aj, 134, 1298

\bibitem[Carrera et al.(2013)]{2013MNRAS.434.1681C} Carrera, R., Pancino, E., Gallart, C., \& del Pino, A.\ 2013, \mnras, 434, 1681

\bibitem[Carretta et al.(2007)]{2007A&A...464..967C} Carretta, E., Bragaglia, A., Gratton, R.~G., et al.\ 2007, \aap, 464, 967

\bibitem[Carretta et al.(2009a)]{2009A&A...505..117C} Carretta, E., Bragaglia, A., Gratton, R.~G., et al.\ 2009a, \aap, 505, 117

\bibitem[Carretta et al.(2009b)]{2009A&A...505..139C} Carretta, E., Bragaglia, A., Gratton, R., \& Lucatello, S.\ 2009b, \aap, 505, 139

\bibitem[Carretta et al.(2009c)]{2009A&A...508..695C} Carretta, E., Bragaglia, A., Gratton, R., D'Orazi, V., \& Lucatello, S.\ 2009c, \aap, 508, 695

\bibitem[Carretta et al.(2010a)]{2010A&A...520A..95C} Carretta, E., Bragaglia, A., Gratton, R.~G., et al.\ 2010a, \aap, 520, A95

\bibitem[Carretta et al.(2010b)]{2010ApJ...722L...1C} Carretta, E., Gratton, R.~G., Lucatello, S., et al.\ 2010b, \apjl, 722, L1

\bibitem[Carretta et al.(2011)]{2011A&A...533A..69C} Carretta, E., Lucatello, S., Gratton, R.~G., Bragaglia, A., \& D'Orazi, V.\ 2011, \aap, 533, A69

\bibitem[Carretta et al.(2012)]{2012A&A...543A.117C} Carretta, E., D'Orazi, V., Gratton, R.~G., \& Lucatello, S.\ 2012, \aap, 543, A117

\bibitem[Carretta et al.(2013)]{2013ApJ...769...40C} Carretta, E., Gratton, R.~G., Bragaglia, A., et al.\ 2013, \apj, 769, 40

\bibitem[Carretta et al.(2014)]{2014A&A...564A..60C} Carretta, E., Bragaglia, A., Gratton, R.~G., et al.\ 2014, \aap, 564, A60

\bibitem[Carretta(2014)]{2014ApJ...795L..28C} Carretta, E.\ 2014, \apjl, 795, L28

\bibitem[Carretta(2015)]{2015ApJ...810..148C} Carretta, E.\ 2015, \apj, 810, 148

\bibitem[Castelli \& Kurucz(2004)]{2004astro.ph..5087C} Castelli, F., \& Kurucz, R.~L.\ 2004, arXiv:astro-ph/0405087

\bibitem[Chen \& Chen(2010)]{2010ApJ...721.1790C} Chen, C.~W., \& Chen, W.~P.\ 2010, \apj, 721, 1790

\bibitem[Cohen(1978)]{1978ApJ...223..487C} Cohen, J.~G.\ 1978, \apj, 223,
487

\bibitem[Cohen(2004)]{2004AJ....127.1545C} Cohen, J.~G.\ 2004, \aj, 127, 1545 

\bibitem[Cohen \& Mel{\'e}ndez(2005)]{2005AJ....129..303C} Cohen, J.~G., \& Mel{\'e}ndez, J.\ 2005, \aj, 129, 303

\bibitem[Cohen \& Kirby(2012)]{2012ApJ...760...86C} Cohen, J.~G., \& Kirby, E.~N.\ 2012, \apj, 760, 86

\bibitem[Cole et al.(2004)]{2004MNRAS.347..367C} Cole, A.~A., Smecker-Hane, T.~A., Tolstoy, E., Bosler, T.~L., \& Gallagher, J.~S.\ 2004, \mnras, 347, 367

\bibitem[Cordero et al.(2014)]{2014ApJ...780...94C} Cordero, M.~J., Pilachowski, C.~A., Johnson, C.~I., et al.\ 2014, \apj, 780, 94

\bibitem[C{\^o}t{\'e} et al.(1995)]{1995ApJ...454..788C} C{\^o}t{\'e}, P., Welch, D.~L., Fischer, P., \& Gebhardt, K.\ 1995, \apj, 454, 788

\bibitem[Cottrell \& Da Costa(1981)]{1981ApJ...245L..79C} Cottrell, P.~L., \& Da Costa, G.~S.\ 1981, \apjl, 245, L79

\bibitem[D'Antona et al.(2002)]{2002A&A...395...69D} D'Antona, F., Caloi, V., Montalb{\'a}n, J., Ventura, P., \& Gratton, R.\ 2002, \aap, 395, 69

\bibitem[D'Antona et al.(2010)]{2010MNRAS.405.2295D} D'Antona, F., Caloi, V., \& Ventura, P.\ 2010, \mnras, 405, 2295

\bibitem[D'Antona et al.(2016)]{2016MNRAS.458.2122D} D'Antona, F., Vesperini, E., D'Ercole, A., et al.\ 2016, \mnras, 458, 2122 

\bibitem[D'Cruz et al.(1996)]{1996ApJ...466..359D} D'Cruz, N.~L., Dorman, B., Rood, R.~T., \& O'Connell, R.~W.\ 1996, \apj, 466, 359

\bibitem[D'Orazi et al.(2010)]{2010ApJ...719L.213D} D'Orazi, V., Gratton, R., Lucatello, S., et al.\ 2010, \apjl, 719, L213

\bibitem[Da Costa et al.(2009)]{2009ApJ...705.1481D} Da Costa, G.~S., Held, E.~V., Saviane, I., \& Gullieuszik, M.\ 2009, \apj, 705, 1481

\bibitem[Da Costa et al.(2013)]{2013ApJ...769....8D} Da Costa, G.~S., Norris, J.~E., \& Yong, D.\ 2013, \apj, 769, 8

\bibitem[Da Costa(2016a)]{2016IAUS..317..110D} Da Costa, G.~S.\ 2016a, The General Assembly of Galaxy Halos: Structure, Origin and Evolution, 317, 110

\bibitem[Da Costa(2016b)]{2016MNRAS.455..199D} Da Costa, G.~S.\ 2016b, \mnras, 455, 199

\bibitem[Davidge(2000)]{2000AJ....120.1853D} Davidge, T.~J.\ 2000, \aj, 120, 1853

\bibitem[Davies(2015)]{2015ebss.book..203D} Davies, M.~B.\ 2015, Ecology of Blue Straggler Stars, 203

\bibitem[Dieball et al.(2009)]{2009MNRAS.394L..56D} Dieball, A., Knigge, C., Maccarone, T.~J., et al.\ 2009, \mnras, 394, L56

\bibitem[Dolphin(2000)]{2000PASP..112.1383D} Dolphin, A.~E.\ 2000, \pasp, 112, 1383

\bibitem[Dotter et al.(2008)]{2008ApJS..178...89D} Dotter, A., Chaboyer, B., Jevremovi{\'c}, D., et al.\ 2008, \apjs, 178, 89-101

\bibitem[Dupree et al.(2011)]{2011ApJ...728..155D} Dupree, A.~K., Strader, J., \& Smith, G.~H.\ 2011, \apj, 728, 155

\bibitem[Dupree et al.(2016)]{2016ApJ...821L...7D} Dupree, A.~K., Avrett, E.~H., \& Kurucz, R.~L.\ 2016, \apjl, 821, L7

\bibitem[Ferraro et al.(2009)]{2009Natur.462..483F} Ferraro, F.~R., Dalessandro, E., Mucciarelli, A., et al.\ 2009, \nat, 462, 483

\bibitem[Garc{\'{\i}}a-Hern{\'a}ndez et al.(2015)]{2015ApJ...815L...4G} Garc{\'{\i}}a-Hern{\'a}ndez, D.~A., M{\'e}sz{\'a}ros, S., Monelli, M., et al.\ 2015, \apjl, 815, L4

\bibitem[Gavagnin et al.(2016)]{2016MNRAS.461.1276G} Gavagnin, E., Mapelli, M., \& Lake, G.\ 2016, \mnras, 461, 1276

\bibitem[Girardi et al.(2008)]{2008PASP..120..583G} Girardi, L., Dalcanton, J., Williams, B., et al.\ 2008, \pasp, 120, 583

\bibitem[Gosling et al.(2009)]{2009MNRAS.394.2247G} Gosling, A.~J., Bandyopadhyay, R.~M., \& Blundell, K.~M.\ 2009, \mnras, 394, 2247

\bibitem[Gratton et al.(2001)]{2001A&A...369...87G} Gratton, R.~G., Bonifacio, P., Bragaglia, A., et al.\ 2001, \aap, 369, 87

\bibitem[Gratton et al.(2004)]{2004ARA&A..42..385G} Gratton, R., Sneden, C., \& Carretta, E.\ 2004, \araa, 42, 385

\bibitem[Gratton et al.(2010)]{2010A&A...522A..77G} Gratton, R.~G., D'Orazi, V., Bragaglia, A., Carretta, E., \& Lucatello, S.\ 2010, \aap, 522, A77

\bibitem[Greggio \& Renzini(1990)]{1990ApJ...364...35G} Greggio, L., \& Renzini, A.\ 1990, \apj, 364, 35

\bibitem[Grillmair et al.(1995)]{1995AJ....109.2553G} Grillmair, C.~J., Freeman, K.~C., Irwin, M., \& Quinn, P.~J.\ 1995, \aj, 109, 2553

\bibitem[Grundahl et al.(1998)]{1998ApJ...500L.179G} Grundahl, F., VandenBerg, D.~A., \& Andersen, M.~I.\ 1998, \apjl, 500, L179

\bibitem[Grundahl et al.(1999)]{1999ApJ...524..242G} Grundahl, F., Catelan, M., Landsman, W.~B., Stetson, P.~B., \& Andersen, M.~I.\ 1999, \apj, 524, 242

\bibitem[Han et al.(2015)]{2015ApJ...813L..43H} Han, S.-I., Lim, D., Seo, H., \& Lee, Y.-W.\ 2015, \apjl, 813, L43

\bibitem[Harris et al.(1976)]{1976ApJS...31...13H} Harris, W.~E., Racine, R., \& de Roux, J.\ 1976, \apjs, 31, 13

\bibitem[Harris(1996)]{1996AJ....112.1487H} Harris, W.~E.\ 1996, \aj, 112, 1487

\bibitem[Idiart et al.(1997)]{1997AJ....113.1066I} Idiart, T.~P., Thevenin, F., \& de Freitas Pacheco, J.~A.\ 1997, \aj, 113, 1066

\bibitem[Ivans et al.(1999)]{1999AJ....118.1273I} Ivans, I.~I., Sneden, C., Kraft, R.~P., et al.\ 1999, \aj, 118, 1273

\bibitem[Ivans et al.(2001)]{2001AJ....122.1438I} Ivans, I.~I., Kraft, R.~P., Sneden, C., et al.\ 2001, \aj, 122, 1438

\bibitem[James et al.(2004)]{2004A&A...427..825J} James, G., Fran{\c c}ois, P., Bonifacio, P., et al.\ 2004, \aap, 427, 825

\bibitem[Johnson \& Pilachowski(2010)]{2010ApJ...722.1373J} Johnson, C.~I., \& Pilachowski, C.~A.\ 2010, \apj, 722, 1373

\bibitem[Johnson \& Pilachowski(2012)]{2012ApJ...754L..38J} Johnson, C.~I., \& Pilachowski, C.~A.\ 2012, \apjl, 754, L38

\bibitem[Johnson et al.(2013)]{2013ApJ...765..157J} Johnson, C.~I., Rich, R.~M., Kobayashi, C., et al.\ 2013, \apj, 765, 157

\bibitem[Johnson et al.(2015a)]{2015AJ....150...63J} Johnson, C.~I., Rich, R.~M., Pilachowski, C.~A., et al.\ 2015a, \aj, 150, 63

\bibitem[Johnson et al.(2015b)]{2015AJ....149...71J} Johnson, C.~I., McDonald, I., Pilachowski, C.~A., et al.\ 2015b, \aj, 149, 71

\bibitem[Kacharov et al.(2014)]{2014A&A...567A..69K} Kacharov, N., Bianchini, P., Koch, A., et al.\ 2014, \aap, 567, A69

\bibitem[Kappeler et al.(1989)]{1989RPPh...52..945K} Kappeler, F., Beer,
H., \& Wisshak, K.\ 1989, Reports on Progress in Physics, 52, 945

\bibitem[Kimmig et al.(2015)]{2015AJ....149...53K} Kimmig, B., Seth, A., Ivans, I.~I., et al.\ 2015, \aj, 149, 53

\bibitem[Kraft et al.(1997)]{1997AJ....113..279K} Kraft, R.~P., Sneden, C., Smith, G.~H., et al.\ 1997, \aj, 113, 279

\bibitem[Kunder et al.(2012)]{2012AJ....143...57K} Kunder, A., Koch, A., Rich, R.~M., et al.\ 2012, \aj, 143, 57

\bibitem[Kurtz \& Mink(1998)]{1998PASP..110..934K} Kurtz, M.~J., \& Mink, D.~J.\ 1998, \pasp, 110, 934

\bibitem[Kuzma et al.(2016)]{2016MNRAS.461.3639K} Kuzma, P.~B., Da Costa, G.~S., Mackey, A.~D., \& Roderick, T.~A.\ 2016, \mnras, 461, 3639 

\bibitem[Lane et al.(2009)]{2009MNRAS.400..917L} Lane, R.~R., Kiss, L.~L., Lewis, G.~F., et al.\ 2009, \mnras, 400, 917

\bibitem[Lane et al.(2010a)]{2010MNRAS.401.2521L} Lane, R.~R., Kiss, L.~L., Lewis, G.~F., et al.\ 2010a, \mnras, 401, 2521

\bibitem[Lane et al.(2010b)]{2010MNRAS.406.2732L} Lane, R.~R., Kiss, L.~L., Lewis, G.~F., et al.\ 2010b, \mnras, 406, 2732

\bibitem[Langer et al.(1993)]{1993PASP..105..301L} Langer, G.~E., Hoffman, R., \& Sneden, C.\ 1993, \pasp, 105, 301

\bibitem[Langer et al.(1997)]{1997PASP..109..244L} Langer, G.~E., Hoffman, R.~E., \& Zaidins, C.~S.\ 1997, \pasp, 109, 244

\bibitem[Lapenna et al.(2014)]{2014ApJ...797..124L} Lapenna, E., Mucciarelli, A., Lanzoni, B., et al.\ 2014, \apj, 797, 124

\bibitem[Lapenna et al.(2016)]{2016ApJ...826L...1L} Lapenna, E., Lardo, C., Mucciarelli, A., et al.\ 2016, \apjl, 826, L1

\bibitem[Lardo et al.(2011)]{2011A&A...525A.114L} Lardo, C., Bellazzini, M., Pancino, E., et al.\ 2011, \aap, 525, A114

\bibitem[Lardo et al.(2015)]{2015A&A...573A.115L} Lardo, C., Pancino, E., Bellazzini, M., et al.\ 2015, \aap, 573, A115

\bibitem[Lardo et al.(2016)]{2016MNRAS.457...51L} Lardo, C., Mucciarelli, A., \& Bastian, N.\ 2016, \mnras, 457, 51

\bibitem[Latour et al.(2014)]{2014ApJ...795..106L} Latour, M., Randall, S.~K., Fontaine, G., et al.\ 2014, \apj, 795, 106

\bibitem[Law \& Majewski(2010)]{2010ApJ...718.1128L} Law, D.~R., \& Majewski, S.~R.\ 2010, \apj, 718, 1128

\bibitem[Lawler et al.(2001)]{2001ApJ...563.1075L} Lawler, J.~E., Wickliffe, M.~E., den Hartog, E.~A., \& Sneden, C.\ 2001, \apj, 563, 1075

\bibitem[Lee(2015)]{2015ApJS..219....7L} Lee, J.-W.\ 2015, \apjs, 219, 7

\bibitem[Lee(2016)]{2016arXiv160808297L} Lee, J.-W.\ 2016, arXiv:1608.08297

\bibitem[Lim et al.(2015)]{2015ApJS..216...19L} Lim, D., Han, S.-I., Lee, Y.-W., et al.\ 2015, \apjs, 216, 19

\bibitem[Lind et al.(2012)]{2012MNRAS.427...50L} Lind, K., Bergemann, M., \& Asplund, M.\ 2012, \mnras, 427, 50

\bibitem[Mackey \& van den Bergh(2005)]{2005MNRAS.360..631M} Mackey, A.~D., \& van den Bergh, S.\ 2005, \mnras, 360, 631

\bibitem[MacLean et al.(2016)]{2016MNRAS.460L..69M} MacLean, B.~T., Campbell, S.~W., De Silva, G.~M., et al.\ 2016, \mnras, 460, L69

\bibitem[Mallia(1978)]{1978A&A....70..115M} Mallia, E.~A.\ 1978, \aap, 70, 115

\bibitem[Marino et al.(2009)]{2009A&A...505.1099M} Marino, A.~F., Milone, A.~P., Piotto, G., et al.\ 2009, \aap, 505, 1099

\bibitem[Marino et al.(2011a)]{2011ApJ...731...64M} Marino, A.~F., Milone, A.~P., Piotto, G., et al.\ 2011a, \apj, 731, 64

\bibitem[Marino et al.(2011b)]{2011A&A...532A...8M} Marino, A.~F., Sneden, C., Kraft, R.~P., et al.\ 2011b, \aap, 532, A8

\bibitem[Marino et al.(2014a)]{2014MNRAS.437.1609M} Marino, A.~F., Milone, A.~P., Przybilla, N., et al.\ 2014a, \mnras, 437, 1609

\bibitem[Marino et al.(2014b)]{2014MNRAS.442.3044M} Marino, A.~F., Milone, A.~P., Yong, D., et al.\ 2014b, \mnras, 442, 3044

\bibitem[Marino et al.(2015)]{2015MNRAS.450..815M} Marino, A.~F., Milone, A.~P., Karakas, A.~I., et al.\ 2015, \mnras, 450, 815

\bibitem[Marino et al.(2016)]{2016MNRAS.459..610M} Marino, A.~F., Milone, A.~P., Casagrande, L., et al.\ 2016, \mnras, 459, 610

\bibitem[Massari et al.(2014)]{2014ApJ...795...22M} Massari, D., Mucciarelli, A., Ferraro, F.~R., et al.\ 2014, \apj, 795, 22

\bibitem[Mateo et al.(2012)]{2012SPIE.8446E..4YM} Mateo, M., Bailey, J.~I., Crane, J., et al.\ 2012, \procspie, 8446, 84464Y

\bibitem[Mauro et al.(2014)]{2014A&A...563A..76M} Mauro, F., Moni Bidin, C., Geisler, D., et al.\ 2014, \aap, 563, A76

\bibitem[McWilliam(1997)]{1997ARA&A..35..503M} McWilliam, A.\ 1997, \araa, 35, 503

\bibitem[McWilliam et al.(2013)]{2013ApJ...778..149M} McWilliam, A., Wallerstein, G., \& Mottini, M.\ 2013, \apj, 778, 149

\bibitem[M{\'e}sz{\'a}ros et al.(2015)]{2015AJ....149..153M} M{\'e}sz{\'a}ros, S., Martell, S.~L., Shetrone, M., et al.\ 2015, \aj, 149, 153

\bibitem[Milone et al.(2012)]{2012A&A...540A..16M} Milone, A.~P., Piotto, G., Bedin, L.~R., et al.\ 2012, \aap, 540, A16

\bibitem[Milone et al.(2013)]{2013ApJ...767..120M} Milone, A.~P., Marino, A.~F., Piotto, G., et al.\ 2013, \apj, 767, 120

\bibitem[Milone et al.(2015a)]{2015MNRAS.447..927M} Milone, A.~P., Marino, A.~F., Piotto, G., et al.\ 2015a, \mnras, 447, 927

\bibitem[Milone et al.(2015b)]{2015ApJ...808...51M} Milone, A.~P., Marino, A.~F., Piotto, G., et al.\ 2015b, \apj, 808, 51

\bibitem[Moehler et al.(2000)]{2000A&A...360..120M} Moehler, S., Sweigart, A.~V., Landsman, W.~B., \& Heber, U.\ 2000, \aap, 360, 120

\bibitem[Moehler et al.(2007)]{2007A&A...475L...5M} Moehler, S., Dreizler, S., Lanz, T., et al.\ 2007, \aap, 475, L5

\bibitem[Moehler et al.(2011)]{2011A&A...526A.136M} Moehler, S., Dreizler, S., Lanz, T., et al.\ 2011, \aap, 526, A136

\bibitem[Momany et al.(2002)]{2002ApJ...576L..65M} Momany, Y., Piotto, G., Recio-Blanco, A., et al.\ 2002, \apjl, 576, L65

\bibitem[Momany et al.(2004)]{2004A&A...420..605M} Momany, Y., Bedin, L.~R., Cassisi, S., et al.\ 2004, \aap, 420, 605

\bibitem[Mucciarelli et al.(2008)]{2008AJ....136..375M} Mucciarelli, A., Carretta, E., Origlia, L., \& Ferraro, F.~R.\ 2008, \aj, 136, 375

\bibitem[Mucciarelli et al.(2012)]{2012MNRAS.426.2889M} Mucciarelli, A., Bellazzini, M., Ibata, R., et al.\ 2012, \mnras, 426, 2889

\bibitem[Mucciarelli et al.(2014)]{2014ApJ...786...14M} Mucciarelli, A., Lovisi, L., Lanzoni, B., \& Ferraro, F.~R.\ 2014, \apj, 786, 14

\bibitem[Mucciarelli et al.(2015a)]{2015ApJ...809..128M} Mucciarelli, A., Lapenna, E., Massari, D., et al.\ 2015a, \apj, 809, 128

\bibitem[Mucciarelli et al.(2015b)]{2015ApJ...801...69M} Mucciarelli, A., Lapenna, E., Massari, D., Ferraro, F.~R., \& Lanzoni, B.\ 2015b, \apj, 801, 69

\bibitem[Mucciarelli et al.(2015c)]{2015ApJ...801...68M} Mucciarelli, A., Bellazzini, M., Merle, T., et al.\ 2015c, \apj, 801, 68

\bibitem[Nataf et al.(2013)]{2013ApJ...769...88N} Nataf, D.~M., Gould, A., Fouqu{\'e}, P., et al.\ 2013, \apj, 769, 88

\bibitem[Nataf et al.(2016)]{2016MNRAS.456.2692N} Nataf, D.~M., Gonzalez, O.~A., Casagrande, L., et al.\ 2016, \mnras, 456, 2692

\bibitem[Navin et al.(2015)]{2015MNRAS.453..531N} Navin, C.~A., Martell, S.~L., \& Zucker, D.~B.\ 2015, \mnras, 453, 531

\bibitem[Navin et al.(2016)]{2016arXiv160606430N} Navin, C.~A., Martell, S.~L., \& Zucker, D.~B.\ 2016, arXiv:1606.06430

\bibitem[Ness et al.(2013a)]{2013MNRAS.432.2092N} Ness, M., Freeman, K., Athanassoula, E., et al.\ 2013a, \mnras, 432, 2092

\bibitem[Ness et al.(2013b)]{2013MNRAS.430..836N} Ness, M., Freeman, K., Athanassoula, E., et al.\ 2013b, \mnras, 430, 836

\bibitem[Nomoto et al.(2006)]{2006NuPhA.777..424N} Nomoto, K., Tominaga, N., Umeda, H., Kobayashi, C., \& Maeda, K.\ 2006, Nuclear Physics A, 777, 424

\bibitem[Norris et al.(1981)]{1981ApJ...244..205N} Norris, J., Cottrell, P.~L., Freeman, K.~C., \& Da Costa, G.~S.\ 1981, \apj, 244, 205

\bibitem[Norris \& Da Costa(1995)]{1995ApJ...447..680N} Norris, J.~E., \& Da Costa, G.~S.\ 1995, \apj, 447, 680

\bibitem[Norris et al.(1996)]{1996ApJ...462..241N} Norris, J.~E., Freeman, K.~C., \& Mighell, K.~J.\ 1996, \apj, 462, 241

\bibitem[Olszewski et al.(1991)]{1991AJ....101..515O} Olszewski, E.~W., Schommer, R.~A., Suntzeff, N.~B., \& Harris, H.~C.\ 1991, \aj, 101, 515

\bibitem[Olszewski et al.(2009)]{2009AJ....138.1570O} Olszewski, E.~W., Saha, A., Knezek, P., et al.\ 2009, \aj, 138, 1570

\bibitem[Origlia et al.(2003)]{2003ApJ...591..916O} Origlia, L., Ferraro, F.~R., Bellazzini, M., \& Pancino, E.\ 2003, \apj, 591, 916

\bibitem[Origlia et al.(2011)]{2011ApJ...726L..20O} Origlia, L., Rich, R.~M., Ferraro, F.~R., et al.\ 2011, \apjl, 726, L20

\bibitem[Origlia et al.(2013)]{2013ApJ...779L...5O} Origlia, L., Massari, D., Rich, R.~M., et al.\ 2013, \apjl, 779, L5

\bibitem[Pancino et al.(2002)]{2002ApJ...568L.101P} Pancino, E., Pasquini, L., Hill, V., Ferraro, F.~R., \& Bellazzini, M.\ 2002, \apjl, 568, L101

\bibitem[Pasquini et al.(2011)]{2011A&A...531A..35P} Pasquini, L., Mauas, P., K{\"a}ufl, H.~U., \& Cacciari, C.\ 2011, \aap, 531, A35

\bibitem[Peterson(1980)]{1980ApJ...237L..87P} Peterson, R.~C.\ 1980, \apjl,
237, L87

\bibitem[Pietrinferni et al.(2006)]{2006ApJ...642..797P} Pietrinferni, A., Cassisi, S., Salaris, M., \& Castelli, F.\ 2006, \apj, 642, 797

\bibitem[Pilachowski et al.(1982)]{1982ApJ...263..187P} Pilachowski, C., Leep, E.~M., Wallerstein, G., \& Peterson, R.~C.\ 1982, \apj, 263, 187

\bibitem[Pilachowski et al.(1996a)]{1996AJ....112..545P} Pilachowski, C.~A.,
Sneden, C., Kraft, R.~P., \& Langer, G.~E.\ 1996a, \aj, 112, 545

\bibitem[Pilachowski et al.(1996b)]{1996AJ....111.1689P} Pilachowski, C.~A., Sneden, C., \& Kraft, R.~P.\ 1996b, \aj, 111, 1689

\bibitem[Piotto et al.(1999)]{1999AJ....118.1727P} Piotto, G., Zoccali, M., King, I.~R., et al.\ 1999, \aj, 118, 1727

\bibitem[Piotto et al.(2007)]{2007ApJ...661L..53P} Piotto, G., Bedin, L.~R., Anderson, J., et al.\ 2007, \apjl, 661, L53

\bibitem[Piotto et al.(2015)]{2015AJ....149...91P} Piotto, G., Milone, A.~P., Bedin, L.~R., et al.\ 2015, \aj, 149, 91

\bibitem[Plummer(1911)]{1911MNRAS..71..460P} Plummer, H.~C.\ 1911, \mnras, 71, 460

\bibitem[Prantzos et al.(2007)]{2007A&A...470..179P} Prantzos, N., Charbonnel, C., \& Iliadis, C.\ 2007, \aap, 470, 179

\bibitem[Pritzl et al.(2005)]{2005AJ....130.2140P} Pritzl, B.~J., Venn, K.~A., \& Irwin, M.\ 2005, \aj, 130, 2140

\bibitem[Racine(1973)]{1973AJ.....78..180R} Racine, R.\ 1973, \aj, 78, 180

\bibitem[Renzini et al.(2015)]{2015MNRAS.454.4197R} Renzini, A., D'Antona, F., Cassisi, S., et al.\ 2015, \mnras, 454, 4197

\bibitem[Rey et al.(2004)]{2004AJ....127..958R} Rey, S.-C., Lee, Y.-W., Ree, C.~H., et al.\ 2004, \aj, 127, 958

\bibitem[Roederer(2011)]{2011ApJ...732L..17R} Roederer, I.~U.\ 2011, \apjl, 732, L17

\bibitem[Roederer \& Thompson(2015)]{2015MNRAS.449.3889R} Roederer, I.~U., \& Thompson, I.~B.\ 2015, \mnras, 449, 3889

\bibitem[Rosenberg et al.(2004)]{2004ApJ...603..135R} Rosenberg, A., Recio-Blanco, A., \& Garc{\'{\i}}a-Mar{\'{\i}}n, M.\ 2004, \apj, 603, 135

\bibitem[Rutledge et al.(1997)]{1997PASP..109..883R} Rutledge, G.~A., Hesser, J.~E., Stetson, P.~B., et al.\ 1997, \pasp, 109, 883

\bibitem[Salinas \& Strader(2015)]{2015ApJ...809..169S} Salinas, R., \& Strader, J.\ 2015, \apj, 809, 169

\bibitem[Saviane et al.(2012)]{2012A&A...540A..27S} Saviane, I., da Costa, G.~S., Held, E.~V., et al.\ 2012, \aap, 540, A27

\bibitem[Shetrone \& Keane(2000)]{2000AJ....119..840S} Shetrone, M.~D., \& Keane, M.~J.\ 2000, \aj, 119, 840

\bibitem[Simmerer et al.(2003)]{2003AJ....125.2018S} Simmerer, J., Sneden, C., Ivans, I.~I., et al.\ 2003, \aj, 125, 2018

\bibitem[Skrutskie et al.(2006)]{2006AJ....131.1163S} Skrutskie, M.~F.,
Cutri, R.~M., Stiening, R., et al.\ 2006, \aj, 131, 1163

\bibitem[Smith \& Norris(1993)]{1993AJ....105..173S} Smith, G.~H., \& Norris, J.~E.\ 1993, \aj, 105, 173

\bibitem[Smith et al.(2000)]{2000AJ....119.1239S} Smith, V.~V., Suntzeff, N.~B., Cunha, K., et al.\ 2000, \aj, 119, 1239

\bibitem[Sneden(1973)]{1973ApJ...184..839S} Sneden, C.\ 1973, \apj, 184, 839

\bibitem[Sneden et al.(1991)]{1991AJ....102.2001S} Sneden, C., Kraft,
R.~P., Prosser, C.~F., \& Langer, G.~E.\ 1991, \aj, 102, 2001

\bibitem[Sneden et al.(2000)]{2000MmSAI..71..657S} Sneden, C., Ivans, I.~I., \& Kraft, R.~P.\ 2000, \memsai, 71, 657

\bibitem[Sneden et al.(2004)]{2004AJ....127.2162S} Sneden, C., Kraft, R.~P., Guhathakurta, P., Peterson, R.~C., \& Fulbright, J.~P.\ 2004, \aj, 127, 2162

\bibitem[Sneden et al.(2008)]{2008ARA&A..46..241S} Sneden, C., Cowan, J.~J., \& Gallino, R.\ 2008, \araa, 46, 241

\bibitem[Sneden et al.(2014)]{2014ApJS..214...26S} Sneden, C., Lucatello, S., Ram, R.~S., Brooke, J.~S.~A., \& Bernath, P.\ 2014, \apjs, 214, 26

\bibitem[Sosin et al.(1997)]{1997ApJ...480L..35S} Sosin, C., Dorman, B., Djorgovski, S.~G., et al.\ 1997, \apjl, 480, L35

\bibitem[Starkenburg et al.(2010)]{2010A&A...513A..34S} Starkenburg, E., Hill, V., Tolstoy, E., et al.\ 2010, \aap, 513, A34

\bibitem[Suntzeff(1981)]{1981ApJS...47....1S} Suntzeff, N.~B.\ 1981, \apjs, 47, 1

\bibitem[Suntzeff \& Kraft(1996)]{1996AJ....111.1913S} Suntzeff, N.~B., \& Kraft, R.~P.\ 1996, \aj, 111, 1913

\bibitem[Tailo et al.(2015)]{2015Natur.523..318T} Tailo, M., D'Antona, F., Vesperini, E., et al.\ 2015, \nat, 523, 318

\bibitem[Timmes et al.(1995)]{1995ApJS...98..617T} Timmes, F.~X., Woosley, S.~E., \& Weaver, T.~A.\ 1995, \apjs, 98, 617 

\bibitem[Tinsley(1979)]{1979ApJ...229.1046T} Tinsley, B.~M.\ 1979, \apj, 229, 1046

\bibitem[Udalski(2003)]{2003ApJ...590..284U} Udalski, A.\ 2003, \apj, 590, 284

\bibitem[Valcarce \& Catelan(2011)]{2011A&A...533A.120V} Valcarce, A.~A.~R., \& Catelan, M.\ 2011, \aap, 533, A120

\bibitem[Valenti et al.(2007)]{2007AJ....133.1287V} Valenti, E., Ferraro, F.~R., \& Origlia, L.\ 2007, \aj, 133, 1287

\bibitem[V{\'a}squez et al.(2015)]{2015A&A...580A.121V} V{\'a}squez, S., Zoccali, M., Hill, V., et al.\ 2015, \aap, 580, A121

\bibitem[Ventura et al.(2012)]{2012ApJ...761L..30V} Ventura, P., D'Antona, F., Di Criscienzo, M., et al.\ 2012, \apjl, 761, L30

\bibitem[Vesperini et al.(2013)]{2013MNRAS.429.1913V} Vesperini, E., McMillan, S.~L.~W., D'Antona, F., \& D'Ercole, A.\ 2013, \mnras, 429, 1913

\bibitem[Villanova et al.(2012)]{2012ApJ...748...62V} Villanova, S., Geisler, D., Piotto, G., \& Gratton, R.~G.\ 2012, \apj, 748, 62

\bibitem[Villanova et al.(2013)]{2013ApJ...778..186V} Villanova, S., Geisler, D., Carraro, G., Moni Bidin, C., \& Mu{\~n}oz, C.\ 2013, \apj, 778, 186

\bibitem[Walker et al.(2011)]{2011MNRAS.415..643W} Walker, A.~R., Kunder, A.~M., Andreuzzi, G., et al.\ 2011, \mnras, 415, 643

\bibitem[Wang et al.(2016)]{2016A&A...592A..66W} Wang, Y., Primas, F., Charbonnel, C., et al.\ 2016, \aap, 592, A66

\bibitem[White \& Shawl(1987)]{1987ApJ...317..246W} White, R.~E., \& Shawl, S.~J.\ 1987, \apj, 317, 246

\bibitem[Williams et al.(2014)]{2014ApJS..215....9W} Williams, B.~F., Lang, D., Dalcanton, J.~J., et al.\ 2014, \apjs, 215, 9

\bibitem[Woosley \& Weaver(1995)]{1995ApJS..101..181W} Woosley, S.~E., \& Weaver, T.~A.\ 1995, \apjs, 101, 181

\bibitem[Worley \& Cottrell(2010)]{2010MNRAS.406.2504W} Worley, C.~C., \& Cottrell, P.~L.\ 2010, \mnras, 406, 2504

\bibitem[Yong et al.(2005)]{2005A&A...438..875Y} Yong, D., Grundahl, F., Nissen, P.~E., Jensen, H.~R., \& Lambert, D.~L.\ 2005, \aap, 438, 875

\bibitem[Yong \& Grundahl(2008)]{2008ApJ...672L..29Y} Yong, D., \& Grundahl, F.\ 2008, \apjl, 672, L29

\bibitem[Yong et al.(2013)]{2013MNRAS.434.3542Y} Yong, D., Mel{\'e}ndez, J., Grundahl, F., et al.\ 2013, \mnras, 434, 3542

\bibitem[Yong et al.(2014)]{2014MNRAS.441.3396Y} Yong, D., Roederer, I.~U., Grundahl, F., et al.\ 2014, \mnras, 441, 3396

\bibitem[Yong et al.(2016)]{2016MNRAS.460.1846Y} Yong, D., Da Costa, G.~S., \& Norris, J.~E.\ 2016, \mnras, 460, 1846

\bibitem[Zoccali et al.(2008)]{2008A&A...486..177Z} Zoccali, M., Hill, V., Lecureur, A., et al.\ 2008, \aap, 486, 177

\bibitem[Zoccali et al.(2014)]{2014A&A...562A..66Z} Zoccali, M., Gonzalez, O.~A., Vasquez, S., et al.\ 2014, \aap, 562, A66

\end{thebibliography}
\end{document}